\documentclass{article}
\usepackage{amssymb}
\usepackage{amsfonts}
\usepackage{amsmath}
\usepackage{perpage}
\usepackage{epigraph}
\MakePerPage{footnote}
\usepackage[titletoc,title]{appendix}
\usepackage{amsthm}

\theoremstyle{definition}

\newtheorem*{remark*}{Remark}

\begin{document}

\title{The Entropic Dynamics of Quantum Scalar Fields Coupled to Gravity}
\author{Selman Ipek\footnote{sipek@albany.edu}\hspace{.15 cm} and Ariel Caticha\footnote{acaticha@albany.edu} \\
%EndAName
{\small Physics Department, University at Albany-SUNY, Albany, NY 12222, USA.%
}}
\date{}
% Keywords
%\keyword{quantum gravity; quantum field theory; entropy; geometry; Hamiltonian dynamics; quantum foundations}
\maketitle

% Abstract (Do not insert blank lines, i.e. \\) 
\abstract{Entropic dynamics (ED) is a general framework for constructing indeterministic dynamical models based on entropic methods. ED has been used to derive or reconstruct both non-relativistic quantum mechanics and quantum field theory in curved space-time. Here we propose a model for a quantum scalar field propagating in a dynamical space-time. The approach rests on a few key ingredients: (1) Rather than modelling the dynamics of the fields, ED models the dynamics of their probabilities. (2) In accordance with the standard entropic methods of inference the dynamics is dictated by information encoded in constraints. (3) The choice of the physically relevant constraints is dictated by principles of symmetry and invariance. The first such principle imposes the preservation of a symplectic structure which leads to a Hamiltonian formalism with its attendant Poisson brackets and action principle. The second symmetry principle is foliation invariance, which following earlier work by Hojman, Kucha\v{r}, and Teitelboim, is implemented as a requirement of path independence. The result is a hybrid ED model that approaches quantum field theory in one limit and classical general relativity in another, but is not fully described by either. A particularly significant prediction of this ED model is that the coupling of quantum fields to gravity implies violations of the quantum superposition principle.}

\section{Introduction}
Without any empirical matter of fact, and none likely on the near horizon, quantum gravity (QG) research has largely split off into distinct channels, each reflecting a different set of attitudes, and yes, \emph{philosophies} directed towards the problem at hand. (See e.g., \cite{Carney et al 2018} for a recent overview of some feasible experimental proposals.) But this state of affairs should not be entirely surprising. Solving the problem of QG should, after all, entail first addressing the issue of what QG even \emph{is} --- indeed, what is meant by \emph{quantum}? What is meant by \emph{gravity}? Which, if any, elements of Einstein's gravity, or the standard quantum formalism, should be abandoned in the transition to QG?

Pursuant to this, a common view is that the gravitational field itself should, in some way, be quantized. There are several routes to accomplishing this.\footnote{Indeed, highlighting the interpretational difficulties of QG, is the way that notions of \emph{quantum} can vary from model to model. A popular class of theories postulates that the combination of quantum theory with gravity should give way to fundamentally discrete models of space and time, such as in the causal set \cite{Surya 2019} and causal dynamical triangulation \cite{Loll 2019} programs.} Most typical, however, is through some manner of quantization algorithm: extending the existing quantum formalism to the gravitational domain by application of the standard \emph{ad hoc} quantization rules to the \emph{appropriate} gravitational degrees of freedom. Such is the tack taken by the principal QG candidates --- string theory (ST) and loop quantum gravity (LQG).\footnote{To be sure, although there are vast differences between canonical approaches to QG, including LQG, and ST (see e.g., \cite{Butterfield Isham 2001}), the adoption of a quantization algorithm is possibly one of the few points of intersection. For some historical overviews in the development of ST and LQG, see the reviews by S. Mukhi \cite{Mukhi 2011} and C. Rovelli \cite{Rovelli 2011}, respectively.}

But such approaches should be met with some suspicion in the context of QG. One issue is the quantization procedure itself. Take models with constraints, such as general relativity (GR), for instance. Should we solve the constraints then quantize, or quantize then solve the constraints? The two methods are not, in general, the same (see e.g., \cite{Henneaux Teitelboim 1994}). Another issue is choosing which degrees of freedom to quantize in the first place. Historically this process has been guided by tight coordination between theory and experiment, allowing trial and error to supplement an incomplete understanding of the quantization process itself. But absent such help from experiment, a more fundamental understanding of quantization may be necessary to make progress.

The lack of clarity around quantization is made particularly acute in the case of the gravitational field, which plays a dual role in GR as an object with genuine dynamical modes, but one that also serves to establish spatial and temporal relationships. This calls into question the precise nature of the gravitational field. Is it just another field to be quantized, along the lines of the gauge fields of Yang-Mills theories, or is it something else entirely? Canonical approaches to QG, such as LQG, have wagered, with varying degrees of success, that the former is true. But there are various clues (see e.g., \cite{Jacobson 1995}-\cite{Caticha 2019b}) suggesting that gravity is an emergent, statistical phenomenon, not unlike temperature and pressure in statistical physics. Would any physicist quantize a temperature field? Perhaps not, but this is exactly what we might be doing when we quantize the gravitational field.

% Also include other work that seeks to develop probabilistic frameworks that can incorporate gravity and alternative causal structures.
All of this suggests that a different attitude with regards to QG may be in order. In the past couple of decades there has been increased interest (see e.g., \cite{Hall Reginatto 2005}-\cite{Doring Isham 2008b}) in developing approaches to QG that seek to treat gravitational considerations hand-in-hand with a more robust understanding of quantum theory (QT). The goal in these cases, however, is not \emph{just} the development of specific models (although such models must necessarily follow), but the construction of entire frameworks that can readily integrate salient features of gravitational and quantum physics alike. One such approach is afforded by entropic dynamics (ED), which is a general framework for constructing indeterministic dynamical models based on the principles of Bayesian probability and entropic inference. (For comprehensive overview of Bayesian and entropic methods, see e.g., \cite{Caticha 2012}\cite{Jaynes Rosenkrantz 1983}.) 

Among the successes of the ED framework are principled derivations of several aspects of the quantum formalism \cite{Caticha 2009}-\cite{Ipek Caticha 2019}, which have led to many key insights. Foremost among these is the notion of probability itself. In ED, probabilities are \emph{Bayesian} --- that is, there are no classical probabilities or quantum probabilities, there are simply probabilities and they follow the usual rules of probability theory \cite{Cox}. But note that this is not a trivial statement. Taking the structure of probability seriously is, in fact, highly constraining. For instance, once we adopt a probabilistic viewpoint, a natural question that follows is, probabilities of what? Are we dealing with the outcomes of an experimental device, the configurations of fields, or the positions of particles? There is no flexibility here. We must choose; a notion that stands in stark contrast to the usual Copenhagen interpretation. Equally constraining is how probabilities are updated in ED. Indeed, it is not enough to simply declare that $|\Psi|^{2}$ yields a Bayesian probability --- one must also demonstrate that the rules for updating those probabilities are in strict accordance with the Bayesian and entropic methods of inference. In short: ED is a dynamical framework driven by constraints.

%, in the words of Feynman \cite{Feynman 2017}, equip us with new and ``different ideas for guessing." To this end, one possible such approach is afforded by entropic dynamics (ED), which is a general framework for constructing indeterministic dynamical systems based on the principles of Bayesian probability and statistical inference. (For comprehensive overview of Bayesian and entropic methods, see e.g., \cite{Caticha 2012}\cite{Jaynes Rosenkrantz 1983}.)

%Another possible approach is afforded by entropic dynamics (ED), which is a general framework for constructing indeterministic dynamical systems based on the principles of Bayesian probability and statistical inference. (For comprehensive overview of Bayesian and entropic methods, see e.g., \cite{Caticha 2012}\cite{Jaynes Rosenkrantz 1983}.)

One of the essential challenges in ED is therefore the \emph{appropriate} identification of constraints. There have been a series of developments in this regard by appealing to principles of symmetry and invariance, which have rich traditions in both physics \cite{Wigner 1967} and inference \cite{Jefferys 1946}\cite{Jaynes 1968} alike.  Early progress in ED \cite{Caticha 2009}\cite{Caticha 2010a}, aided by insights from Nelson's stochastic mechanics \cite{Nelson 1979}, was made by recognizing that conservation of a suitable energy functional could be used as the main criterion for updating the evolving constraints. This allowed ED to model many important aspects of quantum mechanics, leading eventually to a fully Hamiltonian formalism, with its attendant action principle, a symplectic structure, and Poisson brackets \cite{Bartolomeo et al 2014}. However,  a sharper understanding of the deep role played by isometries and symplectic symmetries in QT (see e.g., \cite{Kibble 1979}-\cite{Ashtekar Schilling 1999}) suggested another path where symplectic and metric structures take a more fundamental place in the ED approach \cite{Caticha 2019c}.\footnote{This has been a crucial insight, for instance, for the development of a model for spin-$1/2$ in ED (see e.g., \cite{Carrara Caticha 2020}).} Issues, such as the single-valued nature of the wave function $\Psi$, or more importantly, the linearity of quantum time evolution are clarified from this perspective as resulting from the marriage of symmetry principles with the probabilistic structure of ED.\footnote{For a discussion the single-valuedness condition in quantum mechanics, see e.g., \cite{Merzbacher 1962}. This condition is particularly problematic in approaches to QT that resemble formal aspects of Nelson's stochastic mechanics (see e.g., \cite{Wallstrom 1994}). See \cite{Caticha 2019c} for how this is resolved in ED.}

% In ED, if our model fails to capture the relevant physical information, we do not question the appropriateness of probability theory or the rules for processing information, but question instead the validity or appropriateness of the constraints.

% The failure of the time evolution to be linear raises several questions about ED, such as, is this a deficiency in ED as a framework?

% In ED, linear of quantum time evolution follows from a specific set of symmetries of constraints. Failure of our current model to have such a linear time evolution is not, however, an indictment of the model per se. The appearance of non-linearities in the coupling to gravity may also suggest that symmetries, constraints, etc., that have proved relevant for a fixed background are simply not correct when the background is dynamical. 

%if the time evolution fails to be linear, we do not question the structure of probability or inference, we question the validity. of the constraints

Building on these developments are the previous efforts by Ipek, Abedi, and Caticha (IAC) \cite{Ipek et al 2017}\cite{Ipek et al 2019} to model the ED of quantum scalar fields on a \emph{fixed} curved background. Of particular interest to us are the manifestly covariant methods developed by IAC, which introduced several novel features to ED. One such contribution pertains to the role of time. In ED, time, or better yet, entropic time is constructed as a scheme to keep track of the accumulation of small changes \cite{Caticha 2010a}. In previous work, in the context of a flat space-time \cite{Caticha 2013}\cite{Ipek Caticha 2014}, it was appropriate to introduce a \emph{global} notion of time, in which all spatial points were updated uniformly. As discussed in  \cite{Ipek et al 2017}\cite{Ipek et al 2019}, however, this assumption must be relaxed in a curved space-time in favor of a \emph{local} notion of entropic time. This raises an important challenge, namely, the construction of an updating scheme that is local in nature.

To this end, the work done by IAC in \cite{Ipek et al 2017}\cite{Ipek et al 2019} synthesized two developments in ED: (I) the adoption of a symplectic structure, together with its accompanying Poisson bracket formalism and (II) an updating scheme that unfolds in local entropic time. The two pieces work in tandem. The former allows ED to marshal the full power of the canonical theory, which, in turn, facilitates the desired local time dynamics. In \cite{Ipek et al 2017}\cite{Ipek et al 2019}, this latter ingredient is itself inspired by the seminal works of Dirac \cite{Dirac 1951}\cite{Dirac Lectures} as well as Hojman \emph{et al.} \cite{Hojman Kuchar Teitelboim 1976}, Kucha\v{r} \cite{Kuchar 1973}, and Teitelboim \cite{Teitelboim 1972}\cite{Teitelboim thesis} (DHKT) in their development of covariant canonical methods in classical field theory.\footnote{These approaches could themselves be traced to the earlier ``many-time" efforts of Weiss \cite{Weiss 1938}, Tomonaga \cite{Tomonaga 1946}, Dirac \cite{Dirac 1932}, and Schwinger \cite{Schwinger 1948 I} in the context of relativistic quantum theory.}

Drawing on the ideas of DHKT, the IAC framework proceeds as follows. A chief concern is the notion of an instant, which is defined by a three-dimensional space-like surface embedded in space-time; plus the fields, probability distributions, and so on, that are defined on these surfaces. It is then possible to slice or \emph{foliate} space-time into a sequence of such space-like surfaces. While the decomposition of space-time into spatial slices, indeed, obscures the local Lorentz symmetry of the full pseudo-Riemannian manifold, manifest covariance can be recovered by requiring that the dynamics be unaffected by the particular choice of foliation. This foliation invariance can itself be implemented by a consistency condition dubbed \emph{path independence} \cite{Kuchar 1973}\cite{Teitelboim 1972}: the evolution of all dynamical quantities from an initial to a final surface must be independent of the choice of intermediate surfaces.

% imposes a set of Poisson brackets and constraints to be satisfied by the generators of the local time dynamics.
% The Poisson brackets themselves are required to mirror the structure of the kinematics of deformations.
The requirement of path independence, as laid out by DHKT, thus imposes certain conditions on the generators of the local time dynamics. Crucially important, however, is not \emph{simply} the existence of such conditions, which merely reflects the constraints of path independence, but that these conditions are of a \emph{universal} nature. Put another way, \emph{any} attempt to formulate a dynamics that unfolds in space-time must mirror this pattern, which is itself reflective of the structure of space-time deformations. Thus ED, which is designed as a dynamical scheme that evolves step-by-step from one instant to the next, can be made manifestly covariant by imposing just such a structure. This results in a relativistic ED where \emph{quantum} fields evolve on a fixed \emph{classical} background, generalizing the pioneering efforts of DHKT which dealt solely with the evolution of \emph{classical} fields.

%From this results an ED that provides a framework for the evolution of \emph{quantum} fields on a fixed classical background, thus generalizing the efforts of DHKT which deal only with the evolution of \emph{classical} fields.

%The resulting ED provides a framework for the evolution of \emph{quantum} fields on a fixed classical background, generalizing the efforts of DHKT, which deal only with the evolution of \emph{classical} fields.

%From this perspective, the covariant ED of IAC generalizes the efforts of DHKT, which dealt with the evolution of \emph{classical} fields, by developing a framework suited for the evolution of quantum fields on a classical background.

%From this perspective, the covariant ED developed by IAC generalizes the efforts of DHKT to the quantum domain. That is, whereas DHKT deal with the evolution of \emph{classical} fields on a fixed \emph{classical} background, IAC 

Here we model the indeterministic dynamics of a scalar field $\chi(x)$ interacting with \emph{classical} gravity. Although there is a growing body of work on modifications of classical GR as a prelude to its eventual quantization, in this paper we pursue a more modest goal: to explore the consequence of a first principles derivation of quantum field theory (QFT) while maintaining classical gravity unmodified. Thus our current work, which expands on \cite{Ipek Caticha 2019}, extends the efforts of IAC in one crucial respect by allowing the background geometry itself to become \emph{dynamical}. Following DHKT, the transition to a dynamical background is not accomplished by altering the conditions of path independence, but rather, it is made by an appropriate choice of variables for describing the evolving geometry. The result is a hybrid ED model that approaches QFT in a fixed background in one limit and approaches classical general relativity in another limit, but is not fully described by either. In particular, the model shares some formal similarities with the so-called Semi-classical Einstein equations (SCEE) (see e.g., \cite{Wald 1994}), but here we model the fluctuations of the quantum fields and \emph{derive} their coupling to classical gravity from first principles without the \emph{ad hoc} arguments typically used to justify the SCEE.

To summarize, in a dynamics based on inference the relevant physical information is supplied through constraints. In the present work the constraints are chosen so as to enforce the symmetry and invariance principles that lie at the foundation of quantum theory and general relativity. More explicitly we impose that the dynamics be such as to preserve the symplectic structure and to enforce path independence which amounts to foliation invariance and local Lorentz invariance.

An important feature of the ED derivation of nonrelativistic quantum mechanics \cite{Caticha 2019c} is its analysis of the superposition principle which led to the recognition that the linearity of quantum mechanics is a consequence of the introduction of Hilbert spaces. After all, this is precisely the reason why Hilbert spaces are introduced in the first place: while in principle they are not needed for the formulation of quantum mechanics, their introduction is nevertheless a very convenient calculational trick because it allows one to make use of the calculational advantages of the linearity they induce. A significant result of our present ED reconstruction of a relativistic QFT coupled to gravity is that the dynamics is fundamentally nonlinear. Not only does this imply violations of the quantum superposition principle, but it brings into question the very reason for Hilbert spaces.

Incidentally, this also makes explicit the fundamental disagreement between the ED approach and other approaches such as the orthodox Copenhagen and the many-worlds interpretation.\footnote{Excellent reviews with extended references to the literature are given in e.g. \cite{Stapp 1972}\cite{Leifer 2014}. The many-worlds interpretation is discussed in \cite{Everett 1957}.} In these interpretations the linear structure of the Hilbert space including the superposition principle is given priority while the probabilistic structure is either a secondary addition designed to how to handle those mysterious physical processes called measurements or, as in the many-worlds interpretation, it is avoided altogether. The unwelcome result is that the dynamical and the probabilistic aspects of quantum theory are essentially incompatible with each other. In contrast, ED resolves these problems by giving priority to the probabilistic structure of QM which relegates Hilbert spaces to play a secondary non-fundamental role. Indeed, in the particular problem discussed here -- quantum fields coupled to dynamical gravity -- the dynamics is intrinsically nonlinear, there is no superposition principle, and therefore no reason for Hilbert spaces.

The outline of the paper is the following. In section \ref{ED_Steps}, we review the ED of short steps. Following this, in section \ref{ST_Notation} we introduce some key notation, which is useful in our development of entropic time in section \ref{ED_time}. Key concepts of space-time deformations and ``embeddability" are introduced in section \ref{Embeddability}. In section \ref{ND_ED} we introduce the canonical formalism in ED, which is a necessary step before we review the condition of path independence developed by DHKT in section \ref{Path Independence}. Section \ref{Canonical Rep} outlines the construction of the local generators. In section \ref{ED_DyEqns} we describe the resulting dynamical equations, while in section \ref{ED_QT} we apply these results to obtain an ostensibly quantum theory. We discuss our results in section \ref{conclusion}.

%%%%%%%%%%%%%%%%%%%%%%%%%%%%%%%%%%%%%%%%%%
\section{Statistical Model for Short Steps}\label{ED_Steps}
We present a short review of the ED of infinitesimal steps in curved space-time, adopting the notations and conventions of \cite{Ipek et al 2017}\cite{Ipek et al 2019}\cite{Ipek Caticha 2019}. Here the object of analysis is a single scalar field $\chi \left( x\right)\equiv \chi_{x} $ that populates space and whose values are posited to be definite, but unknown, and thus amenable to being described by probabilities. An entire field configuration, which we denote $\chi $, lives on a $3$-dimensional space $\sigma $, the points of which are labeled by coordinates $x^{i}$ ($i=1,2,3$). The space $\sigma$ is itself curved and comes equipped with a metric $g_{ij}$ that is currently fixed, but that will later become dynamical. A single field configuration $\chi$ is a point in an $\infty $-dimensional configuration space $\mathcal{C}$. Our uncertainty in the values of this field is therefore quantified by a probability distribution $\rho \lbrack \chi ]$ over $\mathcal{C}$, so that the probability that the field attains a value $\hat{\chi}$ in an infinitesimal region of $\mathcal{C}$ is $\text{Prob}[\chi < \hat{\chi} < \chi+\delta\chi]=\rho[\chi]\, D\chi$, where $D\chi$ is an integration measure over $\mathcal{C}$.

\paragraph*{On microstates--- }
In ED the field distributions $\chi_{x}$ play a singularly special role: they define the ontic state of the system.\footnote{A virtue of the ED approach is that it achieves ontological clarity by insisting on a sharp distinction between its ontic and epistemic elements. A concept is ‘ontic’ when it describes something real that exists independently of any observer. In the words of John Bell ontic variables are beables. A concept is ‘epistemic’ when it is related to the state of knowledge, opinion, or belief of an agent, albeit an ideally rational agent. Examples of epistemic quantities are probabilities, entropies, and wave functions. An important point is that the distinctions ontic/epistemic and objective/subjective are not the same. Probabilities are fully epistemic — they are tools for reasoning with incomplete information — but they can lie anywhere in the spectrum from being completely subjective (two different agents can have different beliefs) to being completely objective. In QM, for example, probabilities are epistemic and objective — there is such a thing as an objectively correct assignment of probabilities. We will say that the wave function $\Psi$, which is fully epistemic and objective, represents a ``physical" state when it represents information about an actual ``physical" situation.} This ontological commitment is in direct contrast with the usual Copenhagen interpretation in which such microscopic values become actualized only through the process of measurement. The Bohmian interpretation shares with ED that fact that in both the fields are ontic but the resemblance ends there; Bohmian wave functions are ontic while ED wave functions are fully epistemic \cite{Caticha 2019c}\cite{Bartolomeo Caticha 2016}. The metric $g_{ij}$, on the other hand, in our approach is a \textit{tool} whose purpose is to measure distances, areas, etc., and to characterize the spatial relations between the physical degrees of freedom, the $\chi_{x}$. While the geometry may later become dynamical, we do \textit{not} interpret this to mean that $g_{ij}$ is itself an ontic variable; it is not.\footnote{A model in which the metric tensor is itself of statistical origin is proposed in \cite{Caticha 2016}.} Put another way, the geometric variables enter much like parameters in a typical statistical model. The value of those parameters are important in guiding the distribution of outcomes. However, unlike the ontic variables, their values are not detected directly, but inferred from an ensemble of measurements.

\paragraph*{Maximum Entropy---}
Our goal is to predict the indeterministic dynamics of the scalar field $\chi $ whose statistical features are captured by a probabilistic model. To this end, we make one major assumption: in ED, the fields follow continuous trajectories such that finite changes can be analyzed as an accumulation of many infinitesimally small ones. Such an assumption allows us to focus our interest on obtaining the probability $P\left[ \chi ^{\prime }|\chi \right] $ of a transition from an initial configuration $\chi $ to a neighboring $\chi ^{\prime }=\chi +\Delta \chi $. This is accomplished via the Maximum Entropy (ME) method by maximizing the entropy functional,\footnote{Here we adopt a view of entropy as a \emph{tool} designed for updating probability distributions from a prior to a posterior. (For an accessible introduction, see e.g., \cite{Caticha 2012}.) Such a perspective establishes entropy, beyond just its role in thermodynamics, as an essential asset to inductive inference writ large.}
\begin{eqnarray}
S\left[ P,Q\right] =-\int D\chi ^{\prime }P\left[ \chi ^{\prime }|\chi %
\right] \log \frac{P\left[ \chi ^{\prime }|\chi \right] }{Q\left[ \chi
^{\prime }|\chi \right] },  \label{6 entropy a}
\end{eqnarray}%
relative to a prior $Q\left[ \chi ^{\prime }|\chi \right] $ and subject to appropriate constraints.

\paragraph*{The prior ---}

We adopt a prior $Q\left[ \chi ^{\prime }|\chi \right] $ that incorporates
the information that the fields change by infinitesimally small amounts, but
is otherwise maximally uninformative. In particular, before the constraints are taken into
account, knowledge of the dynamics at one point does not convey information about the dynamics at another point, i.e. the degrees of freedom are \emph{a priori} uncorrelated.

Such a prior can itself be derived from the principle of maximum entropy. Indeed, maximize
\begin{eqnarray}
S[Q,\mu ]=-\int D\chi ^{\prime }\,Q\left[ \chi ^{\prime }|\chi \right] \log 
\frac{Q\left[ \chi ^{\prime }|\chi \right] }{\mu (\chi ^{\prime })}~,
\label{entropy b}
\end{eqnarray}%
relative to the measure $\mu (\chi ^{\prime })$, which we assume to be
uniform, and subject to appropriate constraints.\footnote{Since we deal with infinitesimally short steps, the prior $Q$ turns out to be quite independent of the background measure $\mu$.} The requirement that the
field undergoes changes that are small and uncorrelated is implemented by
imposing an infinite number of independent constraints, one per spatial point $x$,\\
\begin{eqnarray}
\langle \Delta \chi _{x}^{2}\rangle =\int D\chi ^{\prime }\,Q\left[ \chi
^{\prime }|\chi \right] (\Delta \chi _{x})^{2}=\kappa _{x}\,,
\label{Constraint 1}
\end{eqnarray}%
where $\Delta \chi _{x}=\chi _{x}^{\prime }-\chi _{x}$ and the $\kappa _{x}$
are small quantities. The result of maximizing (\ref{entropy b}) subject to (%
\ref{Constraint 1}) and normalization is a product of Gaussians
\begin{eqnarray}
Q\left[ \chi ^{\prime }|\chi \right] \propto \,\exp -\frac{1}{2}\int
dx\,g_{x}^{1/2}\alpha _{x}\left( \Delta \chi _{x}\right) ^{2}~,  \label{prior}
\end{eqnarray}%
where $\alpha _{x}$ are the Lagrange multipliers associated to each
constraint (\ref{Constraint 1}); the scalar density $g_{x}^{1/2}=\,\left(
\det \,g_{ij}\right) ^{1/2}$ is introduced so that $\alpha _{x}$ is a scalar
field. For notational simplicity we write $dx^{\prime }$ instead of $%
d^{3}x^{\prime }$. To enforce the continuity of the motion we will eventually
take the limit $\kappa _{x}\rightarrow 0$ which amounts to taking $\alpha
_{x}\rightarrow \infty $.

\paragraph*{The global constraint--- }

The motion induced by the prior (\ref{prior}) leads to a rather simple diffusion process in the probabilities, in which the field variables evolve independently of each other. To model a dynamics that exhibits correlations and is capable of demonstrating the full suite of quantum effects, such as the superposition of states, interference, and entanglement, however, we require additional structure. This is accomplished by imposing a \emph{single} additional constraint that is \emph{non-local} in space but local in configuration space, which involves the introduction of a \emph{drift} potential $\phi \lbrack \chi ]$ which is a scalar-valued functional defined over the configuration space $\mathcal{C}$. More explicitly, we impose
\begin{eqnarray}
\langle \Delta \phi \rangle = \int D\chi ^{\prime }\,P\left[ \chi ^{\prime }|\chi \right] \int
dx\,\,\Delta \chi _{x}\frac{\delta \phi \left[ \chi \right] }{\delta \chi
_{x}}=\kappa ^{\prime },  \label{Constraint 2}
\end{eqnarray}%
where we require $\kappa^{\prime}\to 0$. (Note that since $\chi _{x}$ and $\Delta \chi _{x}$ are scalars, in order that (\ref{Constraint 2}) be invariant under coordinate transformations of the surface, the derivative $\delta /\delta \chi _{x}$ must transform as a scalar density.)

Before moving on, we discuss the central role of the drift potential in ED. As a dynamics of probabilities ED brings together three ingredients. First, from an inference perspective the dynamics consists in an updating of probabilities which must obey the rules of entropic inference. This requires information codified into constraints and the drift potential is the function that codifies such information. Second, in order to formulate a dynamical theory in which probabilities are generalized coordinates it is only natural to attempt to identify the canonically conjugate generalized momenta and the corresponding symplectic structure. And third, from a purely geometric perspective the trajectories we seek are curves on the space of probability distributions. To this statistical manifold it is natural to associate its tangent bundle. The fibers of such a bundle are spaces of velocity vectors tangent to all curves. It is also natural to consider the associated cotangent bundle the fibers being the spaces of tangent covectors. As we shall see ED brings together these three separate ingredients by identifying the constraint embodied in the drift potential functional with the momenta conjugate to the probabilities, with the cotangent vectors. More explicitly, we identify the three separate pairs: probability/constraint, coordinate/momentum, point/covector. And as we shall see, eventually these three pairs are also identified with yet a fourth pair: the magnitude/phase of the wave function.

\paragraph*{The transition probability ---}

Next we maximize (\ref{6 entropy a}) subject to (\ref{Constraint 2}) and
normalization. The multiplier $\alpha ^{\prime }$ associated to the global constraint (\ref{Constraint 2}) turns out to have no influence on the dynamics: it can be absorbed into the yet undetermined drift potential $\alpha ^{\prime }\phi \rightarrow \phi $, effectively setting $\alpha ^{\prime }=1$.

The result is a Gaussian transition probability distribution, 
\begin{eqnarray}
P\left[ \chi ^{\prime }|\chi \right] \propto \exp -\frac{1}{2}\int dx\,g_{x}^{1/2}\alpha _{x}\left( \Delta
\chi _{x}-\frac{1}{g_{x}^{1/2}\alpha _{x}}\frac{\delta \phi \left[ \chi %
\right] }{\delta \chi _{x}}\right) ^{2}.  \label{Trans Prob}
\end{eqnarray}%
In previous work \cite{Caticha 2013}\cite{Ipek Caticha 2014}, $\alpha _{x}$ was
chosen to be a spatial constant $\alpha $ to reflect the translational
symmetry of flat space. Such a requirement, however, turns out to be inappropriate in the context of curved space-time. Instead, we follow \cite{Ipek et al 2017}\cite{Ipek et al 2019} in allowing $\alpha _{x}$ to remain a non-uniform spatial scalar. This will be a key element in developing our scheme for a local entropic time.

The form of (\ref{Trans Prob}) allows us to present a generic
change, 
\begin{eqnarray}
\Delta \chi _{x}=\left\langle \Delta \chi _{x}\right\rangle +\Delta w_{x}~,\notag
\end{eqnarray}
as resulting from an expected drift $\left\langle \Delta \chi
_{x}\right\rangle $ plus Gaussian fluctuations $\Delta w_{x}$. Computing the expected short step for $\chi_{x}$ gives
\begin{eqnarray}
\left\langle \Delta \chi _{x}\right\rangle =\frac{1}{g_{x}^{1/2}\,\alpha _{x}%
}\frac{\delta \phi \left[ \chi \right] }{\delta \chi _{x}}~,  \label{Exp Step 1}
\end{eqnarray}%
while the fluctuations $\Delta w_{x}$ satisfy,%
\begin{eqnarray}
\left\langle \Delta w_{x}\right\rangle =0\,,\quad \text{and}\hspace{0.4cm}%
\left\langle \Delta w_{x}\Delta w_{x^{\prime }}\right\rangle =\frac{1}{%
g_{x}^{1/2}\alpha _{x}}\delta _{xx^{\prime }}.  \label{Fluctuations}
\end{eqnarray}%
Thus we see that while the expected step size is of order $\Delta \bar{\chi}_{x}\sim 1/\alpha _{x}$, the fluctuations go as $\Delta w_{x}\sim 1/\alpha _{x}^{1/2}$. Thus, for short steps, i.e. $\alpha_{x}\rightarrow \infty $, the fluctuations overwhelm the drift, resulting in a trajectory that is continuous but not, in general, differentiable. Such a model describes a Brownian motion in the field variables $\chi_{x}$.\footnote{As discussed in \cite{Caticha 2019c}\cite{Bartolomeo Caticha 2016}, with appropriate choices of the multipliers $\alpha$ and $\alpha^{\prime}$ the fields or the particles, as the case might be, can be made to follow paths typical of a Brownian motion or alternatively paths that are smooth and resemble a Bohmian motion. It is remarkable that these different paths at the sub-quantum level all lead to the same Schr\"{o}dinger equation.}

% remember to add section.
\section{Some notation}\label{ST_Notation}
The ED developed here deals with the coupled evolution of a quantum scalar field together with a classical dynamical background. The class of theories that allow such evolving geometries are often called instances of \emph{geometrodynamics}. In a typical geometrodynamics (see e.g., \cite{Dirac 1958}\cite{ADM 1960}\cite{ADM 2008}), the primary object of interest is the evolving three-metric $g_{ij}(x)$, whose dynamics must be suitably constrained so that the time evolution sweeps a four-dimensional space-time with metric $^{4}g_{\mu\nu}$ ($\mu ,\nu ,...$ $=0,1,2,3$). Thus, despite the intrinsic dynamics at play, it is nonetheless appropriate to make reference to the enveloping space-time, if only as formal scaffolding that can be later removed.\footnote{Remarkably, as discussed by Teitelboim \cite{Teitelboim thesis}, although certain aspects of the formalism, indeed, rely crucially on space-time itself, such notions are ultimately absent from the operative equations of geometrodynamics. In view of this, we can rightfully regard three-dimensional space as taking a primary role, with space-time being taken as a secondary construct.}

It is therefore possible to assign coordinates $X^{\mu}$ to the space-time manifold. Furthermore, we deal with a space-time with globally \emph{hyperbolic} topology, admitting a foliation by space-like surfaces $\left\{ \sigma \right\} $. The embedding of such surfaces in space-time is defined by four scalar embedding functions $X^{\mu }\left(x^{i}\right) = X^{\mu}_{x}$. An infinitesimal deformation of the surface $\sigma $ to a neighboring surface $\sigma^{\prime }$ is determined by the deformation vector,
\begin{eqnarray}
\delta \xi ^{\mu }_{x}=\delta \xi ^{\bot }_{x}n^{\mu }_{x}+\delta \xi ^{i}_{x}X_{ix}^{\mu }~.
\label{6 deformation vector}
\end{eqnarray}%
Here we have introduced $n^{\mu }$, which is the unit normal to the surface that is determined by the conditions $n_{\mu }n^{\mu }=-1$ and $n_{\mu }X_{ix}^{\mu }=0$), and where we have introduced $X_{ix}^{\mu} =\partial_{ix}X^{\mu}_{x} $, which are the space-time components of three-vectors tangent to $\sigma$. The normal and tangential components of $\delta \xi ^{\mu }$, also known as the infinitesimal lapse and shift, are collectively denoted $\delta \xi ^{A}_{x}=(\delta \xi
^{\bot }_{x},\delta \xi ^{i}_{x})$ and are given by%
\begin{eqnarray}
\delta \xi _{x}^{\bot }=-n_{\mu x}\delta \xi _{x}^{\mu }\quad \text{and}%
\quad \delta \xi _{x}^{i}=X_{\mu x}^{i}\delta \xi _{x}^{\mu }~,\label{6 deformation components}
\end{eqnarray}
where $X_{\mu x}^{i}=\,^{4}g_{\mu \nu }g^{ij}X_{jx}^{\nu }$. Additionally, a particular deformation is defined by its components $\xi^{A}_{x}$, which allows us to speak unambiguously about applying the same deformations to different surfaces. Consequently, the very same deformation $\xi^{A}_{x}$ applied to different surfaces with distinct normal vectors $n^{\mu}_{x}$ will generally yield different deformation vectors $\xi^{\mu}_{x}$, as per eq.(\ref{6 deformation vector}).

\section{Entropic time}\label{ED_time}
In ED, entropic time is introduced as a tool for keeping track of the accumulation of many short steps. (For additional details on entropic time, see e.g., \cite{Caticha 2012}.) Here we introduce a manifestly covariant notion of entropic time, along the lines of that in \cite{Ipek et al 2017}\cite{Ipek et al 2019}.

\paragraph*{An instant---}
Central to the formulation of entropic time is the notion of an instant which includes two main components. One is kinematic the other informational. The former amounts to specifying a particular space-like surface. The latter consists of specifying the contents of the instant, namely, the information — the relevant probability distributions, drift potentials, geometries, etc.— that are necessary to generate the next instant.

\paragraph*{Ordered instants--- }
Establishing the notion of an instant proves crucial because it supplies the structure necessary to inquire about the field $\chi_{x}$ at a \emph{moment} of time. Equivalently, such a notion allows us to assign a probability distribution $\rho_{\sigma}[\chi]$ corresponding to the informational state of the field $\chi_{x}$ at an instant labeled by the surface $\sigma$.

Dynamics in ED, on the other hand, is constructed step-by-step, as a sequence of instants. Thus it is appropriate to turn our attention to the issue of updating from some distribution $\rho_{\sigma}[\chi]$ at some initial instant, to another distribution $\rho_{\sigma^{\prime}}[\chi]$ at a subsequent instant. Such dynamical information is encoded in the short-step transition probability from eq.(\ref{Trans Prob}), or better yet, the joint probability $ P\left[ \chi ^{\prime },\chi \right]= P\left[ \chi ^{\prime }|\chi \right]\rho_{\sigma}[\chi] $.

Application of the ``sum rule" of probability theory to the joint distribution yields
\begin{eqnarray}
\rho_{\sigma^{\prime}}[\chi^{\prime}] = \int D\chi \, P\left[ \chi ^{\prime }|\chi \right]\rho_{\sigma}[\chi]~.\label{6 CK equation}
\end{eqnarray}
The structure of eq.(\ref{6 CK equation}) is highly suggestive: if we interpret $\rho_{\sigma}[\chi]$ as being an \emph{initial} state, then we can interpret $\rho_{\sigma^{\prime}}[\chi^{\prime}] $ as being \emph{posterior} to it in the sense that it has taken into account the new information captured by the transition probability $P\left[ \chi ^{\prime }|\chi \right]$. Here we take this hint seriously and adopt eq.(\ref{6 CK equation}) as the equation that we seek for updating probabilities.

\paragraph*{Duration--- }
A final aspect of time to be addressed is the duration between instants. In doing so, one must distinguish between two separate issues. On one hand there is a natural notion of time that can be inherited from space-time itself; this being the local proper time $\delta\xi_{x}^{\perp}$ experienced by an observer at the point $x$ (see e.g., \cite{Gourgoulhon 2007}). On the other hand, however, notions of time and duration cannot themselves be divorced from those of dynamics and change. Indeed, the two are closely bound together.

Our strategy can be summarized by Wheeler's maxim [61]: ``time is defined so that motion looks simple." In Newtonian mechanics, for example, time is defined so as to simplify the motion of free particles; the prototype of a clock is a free particle that moves equal distances in equal times. In ED for short steps the dynamics is dominated by fluctuations, eq.(9). Accordingly, the prototype of a clock is provided by the fluctuations of the field. Since the specification of the time interval is achieved by an appropriate choice of multipliers alpha we proceed by setting
\begin{eqnarray}
\alpha _{x}=\frac{1}{ \delta \xi _{x}^{\bot }}\quad \text{so that}\quad
\left\langle \Delta w_{x}\Delta w_{x^{\prime }}\right\rangle =\frac{
\,\delta \xi _{x}^{\bot }}{g_{x}^{1/2}}\delta _{xx^{\prime }}~.
\label{6 Duration}
\end{eqnarray}%
With this the transition probability eq.(\ref{Trans Prob}) resembles a Wiener process, albeit in a rather unfamiliar context involving the propagation of fields on curved space. Equation (\ref{6 Duration}) shows that at each point $x$ time is defined by a local clock so that fluctuations as measured by the variance increase by equal amounts in equal time delta $\xi^{\perp}_{x}$.

\paragraph*{The local-time diffusion equations--- }

The dynamics expressed in integral form by (\ref{6 CK equation}) and (%
\ref{6 Duration}) can be rewritten in differential form. The result is \cite{Ipek et al 2017}\cite{Ipek et al 2019},
\begin{eqnarray}
\delta \rho _{\sigma }\left[ \chi \right] =\int dx\frac{\delta \rho _{\sigma
}\left[ \chi \right] }{\delta \xi _{x}^{\bot }}\delta \xi _{x}^{\bot }=-\int
dx\frac{1}{g_{x}^{1/2}}\frac{\delta }{\delta \chi _{x}}%
\left( \rho _{\sigma }\left[ \chi \right] \frac{\delta \Phi _{\sigma }\left[
\chi \right] }{\delta \chi _{x}}\right) \delta \xi _{x}^{\bot }~, \label{FP b}
\end{eqnarray}%
where we have introduced
\begin{eqnarray}
\Phi_{\sigma }\left[ \chi \right] = \,\phi _{\sigma }\left[ \chi \right]
- \log \rho _{\sigma }^{1/2}\left[ \chi \right]~,\notag%\label{Phi def}
\end{eqnarray}
which we refer to as the \emph{phase} functional.\footnote{A key aspect of ED is that we model the dynamics of probabilities, rather than the corresponding microstates. Therefore, while we do introduce a symplectic structure and fully Hamiltonian formalism in ED, it is for the probability $\rho_{\sigma}$ and its eventual conjugate variable $\Phi_{\sigma}$, not for the ontic field $\chi_{x}$, which, strictly speaking, has no equivalent notion of a conjugate momentum. Nevertheless, an analogous concept can be introduced in ED using the phase functional $\Phi_{\sigma}$ through $P_{x} =  \delta\Phi_{\sigma}/\delta\chi_{x}$. As shown in \cite{Ipek et al 2019}, the \emph{expected values} of $\chi_{x}$ and $P_{x}$ do, in fact, satisfy a Poisson bracket that is very reminiscent of the canonical Poisson bracket relations of classical physics. This suggests identifying $P_{x}$ itself as a type of momentum, which much resembles the notion of momentum familiar from the Hamilton-Jacobi formulation of classical physics.}

For arbitrary choices of the infinitesimal lapse $\delta\xi_{x}^{\perp}$ we obtain an infinite set of local equations, one for each spatial point
\begin{eqnarray}
\frac{\delta \rho _{\sigma }}{\delta \xi _{x}^{\bot }}=-\frac{1}{g_{x}^{1/2}}\frac{%
\delta }{\delta \chi _{x}}\left( \rho _{\sigma }\,\frac{\delta \Phi _{\sigma
}}{\delta \chi _{x}}\right)~. \label{FP equation}
\end{eqnarray}%

To interpret these local equations, consider again the variation given in eq.(\ref{FP b}). In the special case where both surfaces $\sigma $ and $\sigma ^{\prime }$ happen to be flat then $g_{x}^{1/2}=1$ and $\delta \xi _{x}^{\bot }=dt$ are both spatial
\emph{constants} and eq.(\ref{FP b}) becomes equivalent to 
\begin{eqnarray}
\frac{\partial \rho _{t}\left[ \chi \right] }{\partial t}=-\int dx\frac{%
\delta }{\delta \chi _{x}}\left( \rho _{t}\left[ \chi \right] \frac{\delta
\Phi _{t}\left[ \chi \right] }{\delta \chi _{x}}\right) ~.  \label{FP c}
\end{eqnarray}%
We recognize this \cite{Caticha 2013}\cite{Ipek Caticha 2014} as a \emph{diffusion} or Fokker-Planck equation written as a
continuity equation for the flow of probability in configuration space $%
\mathcal{C}$. This suggests identifying the
\begin{eqnarray}
V_{x} = \frac{1}{g_{x}^{1/2}}\frac{\delta\Phi_{\sigma}[\chi]}{\delta\chi_{x}}~\notag%\label{Current vel.}
\end{eqnarray}
that appears in eq.(\ref{FP equation}) as the velocity of the probability current which is valid for curved and flat spaces alike. Accordingly we will refer to (\ref{FP equation}) as the
\textquotedblleft local-time Fokker-Planck\textquotedblright\ equations
(LTFP).

%Returning again to a curved space-time, we can now identify the
%\begin{eqnarray}
%V_{x} = \frac{1}{g_{x}^{1/2}}\frac{\delta\Phi_{\sigma}[\chi]}{\delta\chi_{x}}~\label{Current vel.}
%\end{eqnarray}
%appearing in eq.(\ref{FP equation}) as the current velocity that regulates the flow of probability.

\section{The structure of surface deformations}\label{Embeddability}
Starting with Dirac \cite{Dirac 1951}\cite{Dirac 1958} and developed more fully by Hojman, Kucha\v{r}, and Teitelboim, a chief contribution of the DHKT program was the recognition that covariant dynamical theories had a rich structure that could be traced to the kinematics of surface deformations. Such structure can, however, itself be studied independently of any particular dynamics being considered. We give a brief review of the subject following the presentations of Kucha\v{r} \cite{Kuchar 1973} and Teitelboim \cite{Teitelboim 1972}.

%Starting with Dirac \cite{Dirac 1951}\cite{Dirac 1958} and developed more fully by Hojman, Kucha\v{r}, and Teitelboim, a chief contribution of the DHKT program was uncovering the rich dynamical structure of covariant field theories, including the general theory of relativity itself. A crucial insight in this regard was that a relativistic dynamics is a \emph{constrained} dynamics, whose structure mirrors the kinematics of surface deformations. Such structure can itself be studied independently of any particular dynamics being considered. We give a brief review of the subject following the presentations of Kucha\v{r} \cite{Kuchar 1973} and Teitelboim \cite{Teitelboim 1972}.

%that unfolds in space-time, including the evolution of the background itself, is a \emph{constrained} dynamics, one that mirrors the kinematics of surface deformations. Such structure can itself be studied independently of any particular dynamics being considered. We give a brief review of the subject following the presentations of Kucha\v{r} \cite{Kuchar 1973} and Teitelboim \cite{Teitelboim 1972}.

For simplicity, we consider a generic functional $T\left[ X(x)\right] $ that assigns a real number to every surface defined by the four embedding variables $X^{\mu }(x)$. The variation in the functional $\delta T $ resulting from an arbitrary deformation $\delta \xi _{x}^{A}$ has the form 
\begin{eqnarray}
\delta T=\int dx\,\delta \xi _{x}^{\mu }\frac{\delta T}{\delta \xi _{x}^{\mu
}}=\int dx\,\left( \delta \xi _{x}^{\bot }G_{\bot x}+\delta \xi
_{x}^{i}G_{ix}\right) T~\,,  \label{delta T}
\end{eqnarray}%
where 
\begin{eqnarray}
G_{\bot x}=\frac{\delta }{\delta \xi _{x}^{\bot }}=n_{x}^{\mu }\frac{\delta 
}{\delta X_{x}^{\mu }}\quad \text{and}\quad G_{ix}=\frac{\delta }{\delta \xi
_{x}^{i}}=X_{ix}^{\mu }\frac{\delta }{\delta X_{x}^{\mu }}
\end{eqnarray}%
are the generators of normal and tangential deformations respectively. The generators of deformations $\delta /\delta \xi _{x}^{A}$ form a non-holonomic basis. Thus, unlike the vectors $\delta /\delta X_{x}^{\mu }$, which form a coordinate basis and therefore commute, the generators of deformations have a non-vanishing commutator ``algebra" given by
\begin{eqnarray}
\frac{\delta }{\delta \xi _{x}^{A}}\frac{\delta }{\delta \xi _{x^{\prime}}^{B}}-\frac{\delta }{\delta \xi _{x^{\prime }}^{B}}\frac{\delta }{\delta\xi _{x}^{A}}=\int dx^{\prime \prime }\,\kappa ^{C}{}_{BA}(x^{\prime \prime};x^{\prime },x)\frac{\delta }{\delta \xi _{x^{\prime \prime }}^{C}}
\label{commutator}
\end{eqnarray}%
where $\kappa ^{C}{}_{BA}$ are the \textquotedblleft structure
constants\textquotedblright\ of the \textquotedblleft
group\textquotedblright\ of deformations.

The previous quotes in \textquotedblleft group\textquotedblright\ and \textquotedblleft algebra\textquotedblright\ are a reminder that strictly, the set of deformations do not form a group. The composition of two successive deformations is itself a deformation, of course, but it also depends on the surface to which the first deformation is applied. As we will see below, the ``structure constants" $\kappa ^{C}{}_{BA}$ are not constant, they depend on the metric $g_{ij}$ of the initial surface.\footnote{In a dynamical approach to gravity the metric is itself a functional of the canonical variables. Thus its appearance in the ``algebra" of deformations is in part responsible for the rich structure of geometrodynamics.}

The calculation of $\kappa ^{C}{}_{BA}$ is given in \cite{Kuchar 1973}\cite{Teitelboim 1972}. The key idea is that of \emph{embeddability}, which proceeds as follows. Consider performing two successive infinitesimal deformations $\delta\xi^{A}$ followed by $\delta\eta^{A}$ on an initial surface $\sigma$: $\sigma \overset{\delta \xi }{\rightarrow }\sigma _{1}\overset{\delta \eta }{\rightarrow }\sigma ^{\prime }$. Performing now the very same deformations in the opposite order $\sigma \overset{\delta \eta }{\rightarrow }\sigma _{2}\overset{\delta \xi }{\rightarrow }\sigma ^{\prime \prime }$ yields a final surface $\sigma^{\prime\prime}$ that, in general, differs from $\sigma^{\prime}$. The key point is that since both $\sigma^{\prime}$ and $\sigma^{\prime\prime}$ are embedded in the very same space-time, then there exist be a third deformation $\delta\zeta^{A}$ that relates the two: $\sigma ^{\prime }\overset{\delta \zeta }{\rightarrow }\sigma ^{\prime \prime }$.

As shown by Teitelboim \cite{Teitelboim 1972}, however, the compensating deformation is not at all arbitrary, but can be determined entirely by geometrical arguments:
\begin{eqnarray}
\delta\zeta^{C}_{x^{\prime\prime}} = \int dx\int dx^{\prime}\kappa_{AB}^{C}(x^{\prime\prime};x,x^{\prime})\delta \xi^{A}_{x}\delta\eta^{B}_{x^{\prime}}~,\label{Compensating deformation}
\end{eqnarray}
where the only non-vanishing $\kappa$'s have the form
\begin{subequations}
\begin{eqnarray}
\kappa_{i\perp}^{\perp}(x^{\prime\prime};x,x^{\prime}) &=& -\kappa_{\perp i}^{\perp}(x^{\prime\prime};x^{\prime},x) = -\delta(x^{\prime\prime},x)\partial_{ix^{\prime\prime}}\delta(x^{\prime\prime},x^{\prime})\label{Kappa 1}\\
\kappa_{ij}^{k}(x^{\prime\prime};x,x^{\prime}) &=& -\kappa_{j i}^{k}(x^{\prime\prime};x^{\prime},x)\notag\\
 &=& \delta(x^{\prime\prime},x)\partial_{ix^{\prime\prime}}\delta(x^{\prime\prime},x^{\prime})\delta^{k}_{j}\notag\\
 &&-\delta(x^{\prime\prime},x)\partial_{jx^{\prime\prime}}\delta(x^{\prime\prime},x^{\prime})\delta^{k}_{i}\label{Kappa 2}\\
 \kappa_{\perp\perp}^{i}(x^{\prime\prime};x,x^{\prime}) &=& -\kappa_{\perp \perp}^{i}(x^{\prime\prime};x^{\prime},x) \notag\\
 &=& - g^{ij}(x^{\prime\prime})\delta(x^{\prime\prime},x^{\prime})\partial_{jx^{\prime\prime}}\delta(x^{\prime\prime},x)\notag\\
 &&+ g^{ij}(x^{\prime\prime})\delta(x^{\prime\prime},x)\partial_{jx^{\prime\prime}}\delta(x^{\prime\prime},x^{\prime})\label{Kappa 3}
\end{eqnarray}
\end{subequations}

Identification of the $\kappa$'s implies that the commutator in eq.(\ref{commutator}) satisfies the \textquotedblleft algebra\textquotedblright  \cite{Kuchar 1973}\cite{Teitelboim 1972},
\begin{eqnarray}
\left[G_{Ax},G_{Bx^{\prime}}\right] = \int dx^{\prime\prime}\kappa_{AB}^{C}(x^{\prime\prime};x,x^{\prime})G_{Cx^{\prime\prime}}~,\label{LB Generator Algebra }
\end{eqnarray}
which can now be written more explicitly as
\begin{subequations}
\begin{eqnarray}
\lbrack G_{\bot x},G_{\bot x^{\prime }}] &=&-(g_{x}^{ij}G_{jx}+g_{x^{\prime
}}^{ij}G_{jx^{\prime }})\partial _{ix}\delta (x,x^{\prime })~,  \label{LB1}
\\
\lbrack G_{ix},G_{\bot x^{\prime }}] &=&-G_{\bot x}\partial _{ix}\delta
(x,x^{\prime })~,  \label{LB2} \\
\lbrack G_{ix},G_{jx^{\prime }}] &=&-G_{ix^{\prime }}\,\partial _{jx}\delta
(x,x^{\prime })-G_{jx}\,\partial _{ix}\delta (x,x^{\prime })~,  \label{LB3}
\end{eqnarray}%
\end{subequations}
with all other brackets vanishing.

\section{Entropic geometrodynamics}\label{ND_ED}
In an \textit{entropic} dynamics, evolution is driven by information codified into constraints. An entropic geometrodynamics, it follows, consists of dynamics driven by a specific choice of constraints, which we discuss here. In \cite{Ipek et al 2017}\cite{Ipek et al 2019}, quantum field theory in a curved space-time (QFTCS) was derived under the assumption that the geometry remains fixed. But such assumptions, we know, should break down when one considers states describing a non-negligible concentration of energy and momentum. Thus we must revise our constraints appropriately. A natural way to proceed is thus to allow the geometry itself to take part in the dynamical process: the geometry affects $\rho_{\sigma}[\chi]$ and $\phi_{\sigma}[\chi]$, they then act back on the geometry, and so forth. Our goal here is to make this interplay concrete.

\paragraph*{The canonical updating scheme--- }
A natural question that arises from the above discussion is how to implement the update of the geometry and drift potential $\phi_{\sigma}[\chi]$. Such a task involves two steps. The first is the proper identification of variables for describing the evolving geometry, while the other is the specific manner in which this joint system of variables, including the drift potential $\phi_{\sigma}[\chi]$, is updated. Fortunately, the two challenges can be dealt with quite independently of each other.

In devising a covariant scheme for updating we draw primarily from the work of IAC \cite{Ipek et al 2017}\cite{Ipek et al 2019}. To review briefly, a primary assumption in the IAC approach was the adoption of a canonical framework for governing the coupled dynamics of $\rho_{\sigma}$ and $\phi_{\sigma}$, expressed more conveniently through the transformed variable $\Phi_{\sigma}$. Although this is certainly a strong assumption, it is one that has some justification. On one front, it can be argued that canonical structures seem to have a rather natural place in ED; arising from conservation laws in \cite{Bartolomeo et al 2014}\cite{Ipek Caticha 2014} and alternatively from symmetry considerations \cite{Caticha 2019c}. However, from another perspective entirely, the use of a canonical formalism in ED can also be traced to more pragmatic concerns, as it allows one to borrow from a roster of covariant canonical techniques designed in the context of classical physics (see e.g., \cite{Dirac Lectures}), but deployed for the purposes of ED. These together suggest that we view the canonical setting, and the symplectic symmetries that undergird it, as a central criterion for updating in ED. %But in contrast to the efforts of IAC \cite{Ipek et al 2017}\cite{Ipek et al 2019} in a fixed space-time, here we deal with a fully dynamical background.

\paragraph*{The canonical variables--- }
Crucial to our updating scheme is an appropriate choice of variables. We pursue a conservative approach in which $\rho_{\sigma}$ and $\Phi_{\sigma}$ are packaged together as canonically conjugate variables following the prescription detailed in IAC \cite{Ipek et al 2019}. The nontrivial task of choosing the geometric variables has long been the subject of a lively debate.\footnote{Just to name a few approaches, there are, of course, the original attempts at geometrodynamics from Dirac \cite{Dirac 1958} as well as Arnowitt, Deser, and Misner (ADM) \cite{ADM 1960} that start from the Einstein-Hilbert action and take the metric $g_{ij}$ as the fundamental building block. Somewhat more recently, due in part to the modern success of gauge theories, there has been some interest in taking, not the metric $g_{ij}$, but the Levi-Civita connection $\Gamma^{i}_{jk}$, as the fundamental gravitational object (see e.g., \cite{Ferraris Kijowski 1981}). In a similar spirit is the well-known discovery of Ashtekar \cite{Ashtekar 1986}, which uses triads and spin connections to rewrite GR.} Among these various approaches, however, the pioneering efforts of Hojman, Kucha\v{r}, and Teitelboim (HKT) \cite{Hojman Kuchar Teitelboim 1976} prove to be of special interest, as their work centers a purely canonical approach to the dynamics of geometry. Following HKT, it is possible to take the six components of the metric $g_{ij}$ to be the starting point for a geometrodynamics. The argument for this is grounded in simplicity. If one assumes, as they do, that evolution in local time should mirror the structure of space-time deformations, then the metric appearing on the right hand side of (\ref{Kappa 3}) for the ``structure constant" $\kappa^{i}_{\perp\perp}$ must necessarily be a functional of the canonical variables. Clearly this is most easily satisfied if the metric is itself a canonical variable.

To complete the canonical framework, however, we must also introduce six variables $\pi^{ij}$ that are canonically conjugate to the $g_{ij}$. These variables, the conjugate momenta, are themselves \emph{defined} through the canonical Poisson bracket relations,\footnote{In particular, we do not assume the existence of a Legendre transformation from $\pi^{ij}$ to the ``velocity" of the metric $g_{ij}$.}
\begin{subequations}
\begin{eqnarray}
\big\{ g_{ijx},g_{klx^{\prime}}\big\} &=& \left\{\pi^{ij}_{x},\pi^{kl}_{x^{\prime}}\right \} = 0\\
\left\{g_{ijx},\pi^{kl}_{x^{\prime}}\right \} &=& \delta_{ij}^{kl}\delta(x,x^{\prime}) = \frac{1}{2}\left(\delta^{k}_{i}\delta^{l}_{j}+\delta^{k}_{j}\delta^{l}_{i}\right)\delta(x,x^{\prime})~.\label{6 canonical relations Grav}
\end{eqnarray}
\end{subequations}
That such relations are satisfied is made manifest by writing the Poisson brackets in local coordinates
\begin{eqnarray}
\left\{F,G\right\} &=& \int dx\left(\frac{\delta F}{\delta g_{ijx}}\frac{\delta G}{\delta \pi^{ij}_{x}}-\frac{\delta G}{\delta g_{ijx}}\frac{\delta F}{\delta \pi^{ij}_{x}}\right)\notag\\
&+&\int D\chi\left(\frac{\tilde{\delta} F}{\tilde{\delta}\rho[\chi] }\frac{\tilde{\delta} G}{\tilde{\delta} \Phi[\chi]}-\frac{\tilde{\delta} G}{\tilde{\delta}\rho[\chi] }\frac{\tilde{\delta} F}{\tilde{\delta} \Phi[\chi]}\right)~,\label{6 Poisson brackets}
\end{eqnarray}
where $F$ and $G$ are some arbitrary functionals of the phase space variables. (Note that $\pi^{ij}_{x}$ must be a tensor \emph{density} for the Poisson bracket to transform appropriately under a change of variables of the surface.)

\section{The canonical structure of space-time}\label{Path Independence}
A fully covariant dynamics requires that the updating in local time of all dynamical variables be consistent with the kinematics of surface deformations. Thus, the requirement that the deformed surfaces remain embedded in space-time, which amounts to imposing foliation invariance, translates into a consistency requirement of path independence: if the evolution from an initial instant into a final instant can occur along different paths, then all these paths must lead to the same final values for all dynamical quantities. The approach we adopt for quantum fields coupled to dynamical classical gravity builds on previous work by HKT \cite{Hojman Kuchar Teitelboim 1976} for classical geometrodynamics, and by IAC \cite{Ipek et al 2017}\cite{Ipek et al 2019} for quantum field theory in a non-dynamical space-time.

%Here we lay out a scheme for updating in local time, drawing on the ideas established in ED by IAC in \cite{Ipek et al 2017}\cite{Ipek et al 2019}, which were in turn inspired by the work of DHKT. As laid out by Hojman \emph{et al.} \cite{Hojman Kuchar Teitelboim 1976}, the approach rests on the central premise that local updating is conducted by attributing to each spatial point a set of four Hamiltonian generators; such local Hamiltonians are compactly denoted $H_{Ax}$ with index $A = (\perp , i)$, where $i $ runs over $ 1,2,3$.

%that a set of four Hamiltonian generators be attributed to each spatial point for the purposes of updating in local time; such local Hamiltonians are compactly denoted $H_{Ax}$ with index $A = (\perp , i)$, where $i $ runs over $ 1,2,3$.

Within this scheme the evolution of an arbitrary functional $F$ of the canonical variables is generated by application of a set of local Hamiltonians according to
\begin{eqnarray}
\delta F = \int dx \left \{F, H_{Ax} \right \}\delta\xi^{A}_{x} = \int dx \left (  \left \{F, H_{\perp x} \right \}\delta\xi^{\perp}_{x}+ \left \{F, H_{ix} \right \}\delta\xi^{i}_{x} \right )~,\label{Dyanmical law}
\end{eqnarray}
where parameters $\delta\xi^{A}_{x}$ with $A = (\perp , i = 1,2,3)$ describe an infinitesimal deformation, as per eqns.(\ref{6 deformation vector}) and (\ref{6 deformation components}), and $H_{Ax}$ are the corresponding generators. (Defined in this way, the $H_{Ax}$ turn out to be tensor densities.)

\paragraph*{Path independence--- }
The implementation of path independence \cite{Kuchar 1973}\cite{Teitelboim 1972} then rests on the idea that the Poisson brackets of the generators $H_{Ax}$ form an ``algebra" that closes in the same way, that is, with the same ``structure constants", as the ``algebra" of deformations in eqns.(\ref{LB1})-(\ref{LB3}),
\begin{subequations}
\begin{eqnarray}
\left\{ H_{\bot x},H_{\bot x^{\prime }}\right\} &=&(g_{x}^{ij}H_{jx}+g_{x^{\prime
}}^{ij}H_{jx^{\prime }})\partial _{ix}\delta (x,x^{\prime })~,  \label{6 PB 1}
\\
\left\{ H_{ix},H_{\bot x^{\prime }}\right \} &=&H_{\bot x}\partial _{ix}\delta
(x,x^{\prime })~,  \label{6 PB 2} \\
\left\{ H_{ix},H_{jx^{\prime }}\right \} &=&H_{ix^{\prime }}\,\partial _{jx}\delta
(x,x^{\prime })+H_{jx}\,\partial _{ix}\delta (x,x^{\prime })~.  \label{6 PB 3}
\end{eqnarray}%
\end{subequations}

We conclude this section with two remarks. First, we note that these equations have not been derived. Indeed, imposing (\ref{LB1})-(\ref{LB3}) as strong constraints constitutes the definition of what we mean by imposing consistency between the updating of dynamical variables and the kinematics of surface deformations. But this is not enough. As discussed in detail by Teitelboim \cite{Teitelboim 1972} and HKT \cite{Hojman Kuchar Teitelboim 1976}, one achieves path independence by requiring that the initial values of the canonical variables be restricted to satisfy the \emph{weak} constraints
\begin{eqnarray}
H_{\perp x} \approx 0\quad\text{and}\quad H_{ix} \approx 0 ~.\label{6 Hamiltonian constraints}
\end{eqnarray}
While it is beyond the scope of this paper to explore the full consequences of these weak constraints we merely point out that formally their origin is traced to the fact that $\kappa^{i}_{\perp \perp}$ in eq.(\ref{Kappa 3}) depends on the metric and therefore it is not a true structure ``constant". More physically the constraints represent the fact that the canonical variables $g_{ij}$ and $\pi^{ij}$ are redundant because they represent the true dynamical degrees of freedom plus additional kinematical variables that allow us the freedom to choose space-time coordinates; and, as discussed by Kucha\v{r} \cite{Kuchar 1973}, separating these dynamical and kinematical variables is not an easy task. Furthermore, once satisfied on an initial surface $\sigma$ the dynamics will be such as to preserve these constraints for all subsequent surfaces of the foliation.

%Note that prior to this assumption we could \emph{not}, in fact, interpret the local Hamiltonians $H_{\perp x}$ and $H_{ix}$ as having any particular relationship to space and time. It is only by taking this non-trivial step of relating the conditions of ``integrability" to those of ``embeddability" that this relationship can be established.

We also note that it is only by virtue of relating the conditions of ``integrability" to those of ``embeddability" that we can interpret the role of the local Hamiltonians $H_{\perp x}$ and $H_{ix}$ in relation to space and time. This is a crucial and highly non-trivial step. It is only once this is established that we can interpret $H_{\perp x}$ as a scalar density that is responsible for genuine dynamical evolution and $H_{ix}$ as a vector density that generates spatial diffeomorphisms; the former is called the super-Hamiltonian, while the latter is the so-called super-momentum.

%Moreover, as discussed by Teitelboim \cite{Teitelboim 1980}, in the HKT approach to geometrodynamics weak constraints arise as a set of auxiliary conditions that are \emph{sufficient} for ensuring a path independent evolution and their physical interpretation can be understood using the Dirac-Bergmann approach to constrained systems (see e.g., \cite{Dirac Lectures}). Here we have four constraints per spatial point, each is a so-called \emph{first class} constraint and thus each generates a gauge symmetry. The super-momentum constraint $H_{ix}\approx 0$ generates infinitesimal spatial diffeomorphisms so that an infinitesimal change of coordinates is understood to be a gauge transformation. The super-Hamiltonian constraint, while its interpretation is more subtle, generates a change in the temporal parameterization.

%Note the following: that it is only once we have taken the non-trivial step of relating the conditions of ``integrability" to those of ``embeddability" that we can interpret the role of the local Hamiltonians $H_{\perp x}$ and $H_{ix}$. Having established this, however, it is clear that $H_{ix}$ is a vector density that generates spatial diffeomorphisms, while $H_{\perp x}$ is a scalar density that is responsible for genuine dynamical evolution in local time. This correspondence, moreover, makes clear that we can 

\section{The canonical representation}\label{Canonical Rep}
We now turn our attention to the local Hamiltonian generators $H_{Ax}$, and more specifically, we look to provide explicit expressions for these generators in terms of the canonical variables. This problem was solved in the context of a purely classical geometrodynamics (with or without sources) by HKT in \cite{Hojman Kuchar Teitelboim 1976}. Here we aim to apply their techniques and methodology to a different problem: a geometrodynamics driven by ``quantum" sources. Fortunately, a considerable portion of the HKT formalism can be directly adopted for our purposes.
\subsection*{The super-momentum}
To determine the generators $H_{Ax}[\rho,\Phi;g_{ij},\pi^{ij}]$ it is easiest to begin with the so-called \textit{super-momentum} $H_{ix}[\rho,\Phi;g_{ij},\pi^{ij}]$. This is largely because the function of this generator is well understood: it pushes the canonical variables along the surface they reside on. Since there is no motion ``normal" to the surface, the action of this generator is purely kinematical.

Following HKT \cite{Hojman Kuchar Teitelboim 1976}, consider an infinitesimal tangential deformation such that a point originally labeled by $x^{i}$ is carried to the point previously labeled by $x^{i} + \delta\xi^{i}$. This will induce a corresponding change in any dynamical variables $F$ defined on that surface, $F\to F+\delta F$. This change $\delta F$ can then be computed in two distinct ways, which, of course, must agree. One is by calculating the Lie derivative along $\delta\xi$
\begin{eqnarray}
\delta F = \pounds_{\delta\xi} F~\notag
\end{eqnarray}
and the other is using the super-momentum $H_{ix}$, so that
\begin{eqnarray}
\delta F = \int dx \left\{ F, H_{ix}\right \}\delta\xi^{i}_{x}~.\notag
\end{eqnarray}
\paragraph*{Gravitational super-momentum}
A straightforward example of this is shown for the metric $g_{ij}(x) = g_{ijx}$, which is a rank $(0,2)$ tensor. Its Lie derivative is given by \cite{Schutz}
\begin{eqnarray}
\pounds_{\delta\xi} g_{ij} = \partial_{k}g_{ij}\delta \xi^{k}_{x} + g_{ik}\partial_{j}\delta\xi^{k}_{x} + g_{kj}\partial_{i}\delta\xi^{k}_{x}~.\label{6 Lie derivative metric}
\end{eqnarray}
Alternatively, by using the Poisson brackets we obtain
\begin{eqnarray}
\delta g_{ijx} = \int dx^{\prime}\left\{g_{ijx}, H_{kx^{\prime}}\right \}\delta\xi_{x^{\prime}}^{k} = \int dx^{\prime}\frac{\delta H_{kx^{\prime}}}{\delta\pi^{ij}_{x}}\delta\xi_{x^{\prime}}^{k}~.\notag
\end{eqnarray}
Comparing the two and using the fact that $\delta\xi_{x}^{i}$ is arbitrary yields
\begin{eqnarray}
\frac{\delta H_{kx^{\prime}}}{\delta\pi^{ij}_{x}} = \partial_{kx}g_{ijx}\delta( x,x^{\prime})+ g_{ikx}\partial_{jx}\delta( x,x^{\prime}) + g_{kjx}\partial_{ix}\delta( x,x^{\prime}).\label{6 Tangential deformation metric}
\end{eqnarray}
The delta functions imply that $H_{ix}$ is local in the momentum $\pi^{ij}$.

To fix the dependence on $\pi^{ij}$, in fact, we can use a similar argument as above. Recalling that $\pi^{ij}(x) = \pi^{ij}_{x}$ is a rank $(2,0)$ tensor \emph{density} of weight one, we find that
\begin{eqnarray}
\pounds_{\delta\xi} \pi^{ij} = \partial_{kx}\left(\pi^{ij}\delta\xi^{k}_{x}\right) - \pi^{ik}\partial_{kx}\delta\xi^{j}_{x} - \pi^{kj}\partial_{kx}\delta\xi^{i}_{x}.\label{6 Tangential deformation momentum}
\end{eqnarray}
The same equation can be obtained through the use of a Hamiltonian,
\begin{eqnarray}
\delta\pi^{ij}_{x} = \int dx^{\prime}\left\{\pi^{ij}_{x},H_{kx^{\prime}}\right\}\delta\xi^{k}_{x^{\prime}} = -\int dx^{\prime}\frac{\delta H_{kx^{\prime}}}{\delta g_{ijx}}\delta\xi^{k}_{x^{\prime}} ~,\notag
\end{eqnarray}
so long as $H_{ix}$ satisfies
\begin{eqnarray}
 -\frac{\delta H_{kx^{\prime}}}{\delta g_{ijx}} = \partial_{kx}\left(\pi^{ij}_{x}\delta(x,x^{\prime})\right) - \pi^{il}_{x}\partial_{lx}\delta(x,x^{\prime}) \delta^{j}_{k}- \pi^{lj}_{x}\partial_{lx}\delta(x,x^{\prime})\delta^{i}_{k}~.\label{6 Tangential deformation momentum}
\end{eqnarray}
Integrating these equations for $H_{ix}$ yields
\begin{eqnarray}
H_{ix} =H_{ix}^{G} + \tilde{H}_{ix} ~,\label{6 Super-momentum grav + H}
\end{eqnarray}
where
\begin{eqnarray}
H_{ix}^{G} = -2 \partial_{jx}\left(\pi^{jk}\, g_{ik}\right) + \pi^{jk}\partial_{ix}g_{jk}~, \label{6 Super momentum grav}
\end{eqnarray}
is called the \emph{gravitational} super-momentum, and the functional $\tilde{H}_{ix} = \tilde{H}_{ix}[\rho,\Phi]$ is, at the moment, just an integration ``constant".

Some of these expressions can be simplified by introducing the \emph{covariant} derivative $\nabla_{i}$. For example, using $\nabla_{k} g_{ij} = 0$ we have
\begin{eqnarray}
\pounds_{\delta\xi} g_{ijx} = g_{jk}\nabla_{i}\delta\xi_{x}^{k} + g_{ik}\nabla_{j}\delta\xi_{x}^{k}~\label{6 Tangential deformation metric cov}
\end{eqnarray}
and
\begin{eqnarray}
\pounds_{\delta\xi} \pi^{ij}_{x} = \nabla_{k}\left(\pi^{ij}\delta\xi^{k}_{x}\right) - \pi^{ik}\nabla_{k}\delta\xi^{j}_{x} - \pi^{kj}\nabla_{k}\delta\xi^{i}_{x}~.\label{6 Tangential deformation momentum cov}
\end{eqnarray}
The gravitational super-momentum $H_{ix}^{G}$ also takes the particularly simple form
\begin{eqnarray}
H_{ix}^{G} = -2g_{ik} \nabla_{j}\pi^{jk}= -2\nabla_{j}\left(\pi^{jk}g_{ik}\right) = -2 \nabla_{j}\pi^{j}_{i}~.\label{6 Super momentum grav cov}
\end{eqnarray}

\paragraph*{The ``matter" super-momentum--- }
Next, we turn to the response of the variables $\rho_{\sigma}[\chi]$ and $\Phi_{\sigma}[\chi]$ under a relabeling of the surface coordinates $x^{i} \to x^{i}+\delta\xi^{i}_{x}$. As the coordinates are shifted, so too are the fields defined upon that surface so that
\begin{eqnarray}
\chi_{x} \to \chi_{x} + \pounds_{\delta\xi}\chi_{x}\quad\text{where}\quad \pounds_{\delta\xi}\chi_{x} = \partial_{ix}\chi_{x}\delta\xi^{i}_{x}\notag
\end{eqnarray}
is the Lie derivative for a scalar $\chi_{x}$. This induces a change in the probability, given by
\begin{eqnarray}
\delta\rho_{\sigma}[\chi] \equiv \rho_{\sigma}[\chi + \pounds_{\delta\xi}\chi_{x}] - \rho_{\sigma}[\chi] = \int dx \frac{\delta\rho_{\sigma}[\chi]}{\delta\chi_{x}}\partial_{ix}\chi_{x}\delta\xi^{i}_{x}~.
\end{eqnarray}

Alternatively, this same variation can be computed using the canonical framework
\begin{eqnarray}
\delta\rho_{\sigma}[\chi] = \int dx \left\{\rho_{\sigma}[\chi],H_{ix}\right\}\delta\xi^{i}_{x}~. \label{Tangential deformation rho a}
\end{eqnarray}
If we insert the super-momentum $H_{ix}$ given in eq.(\ref{6 Super-momentum grav + H}), we notice that only the $\tilde{H}_{ix}$ piece will contribute. And so we have
\begin{eqnarray}
\delta\rho_{\sigma}[\chi] = \int dx \left\{\rho_{\sigma}[\chi],\tilde{H}_{ix}\right\}\delta\xi^{i}_{x} = \int dx \frac{\tilde{\delta}\tilde{H}_{ix}}{\tilde{\delta}\Phi_{\sigma}[\chi]}\delta\xi_{x}^{i}~,\label{6 Tangential deformation rho b}
\end{eqnarray}
which for arbitrary $\delta\xi^{i}_{x}$ requires that
\begin{eqnarray}
\frac{\tilde{\delta}\tilde{H}_{ix}}{\tilde{\delta}\Phi_{\sigma}[\chi]} = \frac{\delta\rho_{\sigma}[\chi]}{\delta\chi_{x}}\partial_{ix}\chi_{x}~.\label{6 Tangential deformation rho c}
\end{eqnarray}
A similar argument for $\Phi_{\sigma}[\chi]$ shows that we must also have
\begin{eqnarray}
-\frac{\tilde{\delta}\tilde{H}_{ix}}{\tilde{\delta}\rho_{\sigma}[\chi]} = \frac{\delta\Phi_{\sigma}[\chi]}{\delta\chi_{x}}\partial_{ix}\chi_{x}~,\label{6 Tangential deformation Phi a}
\end{eqnarray}
so that
\begin{eqnarray}
\tilde{H}_{ix} = - \int D\chi \rho_{\sigma}\frac{\delta\Phi_{\sigma}}{\delta\chi_{x}}\partial_{ix}\chi_{x}~.\label{6 Super momentum ensemble}
\end{eqnarray}
Thus the total super-momentum, eq.(\ref{6 Super-momentum grav + H}),
\begin{eqnarray}
 H_{ix} = -2 \nabla_{j}\pi^{j}_{i} - \int D\chi \rho_{\sigma}\frac{\delta\Phi_{\sigma}}{\delta\chi_{x}}\partial_{ix}\chi_{x}\label{6 Super momentum}
\end{eqnarray}
contains two pieces, which we refer to as the gravitational and ``matter" contributions, respectively.\footnote{The division of generators into gravitational and ``matter" pieces established by DHKT is, strictly speaking, an abuse of language. The variables $\rho_{\sigma}$ and $\Phi_{\sigma}$ that constitute ``matter" are more properly understood as describing the statistical state of the material field $\chi_{x}$. Nonetheless, we stick with the convention as a useful shorthand.} Note that there is no gravitational dependence in the ``matter" side, or ``matter" dependence on the gravitational side, that is,
\begin{eqnarray}
H_{ix} = H_{ix}^{G}[g_{ij},\pi^{ij}] + \tilde{H}_{ix}[\rho,\Phi]~.\label{6 Super momentum split}
\end{eqnarray}

By equation (\ref{6 Hamiltonian constraints}) it is, of course, also understood that $H_{ix}$ is subject to the constraint
\begin{eqnarray}
H_{ix}\approx 0~.\label{6 Super momentum constraint}
\end{eqnarray}
Finally, although $H_{ix}$ was obtained, in essence, independently of the Poisson bracket (\ref{6 PB 3}), relating two tangential deformations, it nonetheless satisfies it automatically.\footnote{More technically, $H_{ix}$ is defined by our procedure up to an overall ``constant" vector density $f_{i}(x)$, which is independent of any canonical variables. This, it turns out, is required to vanish by eq.(\ref{6 PB 3}). See e.g., \cite{Hojman Kuchar Teitelboim 1976}.} It is therefore appropriate to view $H_{ix}$ as being completely determined; this is important as it allows us to treat equation (\ref{6 PB 1}) as a set of equations for $H_{\perp x}$ in terms of the known $H_{ix}$.

%An explicit expressions for $H_{ix}$ can be obtained, as we have, by summing the two pieces, which we refer to as the \emph{gravitational} and ``matter" contributions, following the conventions of DHKT. 
\subsection*{The super-Hamiltonian}
We now turn our attention to the generator $H_{\perp x}$ of local time evolution. Following Teitelboim \cite{Teitelboim thesis}, it is useful to decompose $H_{\perp x}$ into two distinct pieces
\begin{eqnarray}
H_{\perp x} = H_{\perp x}^{G}[g_{ij},\pi^{ij}] + \tilde{H}_{\perp x}[\rho,\Phi;g_{ij}, \pi^{ij}]~,\label{6 Super Hamiltonian split a}
\end{eqnarray}
consisting of a gravitational piece $H_{\perp x}^{G}$ depending only on the gravitational variables, and a ``matter" contribution that we suggestively denote by $\tilde{H}_{\perp x}$, called the gravitational and ``matter" super-Hamiltonians, respectively.

As noted by Teitelboim, we make no assumptions in writing $H_{\perp x}$ in this way.\footnote{By construction, we can always identify a piece of the super-Hamiltonian that updates the geometry alone. The ``matter" super-Hamiltonian is then just defined as the difference between the total and gravitational pieces.} Given this splitting, however, we do make the following simplifying assumption: we require the ``matter" super-Hamiltonian $\tilde{H}_{\perp x}$ to be independent of the gravitational momentum $\pi^{ij}$, i.e.,
\begin{eqnarray}
\tilde{H}_{\perp x} = \tilde{H}_{\perp x}[\rho,\Phi;g_{ij}]~\notag
\end{eqnarray}
so that the super-Hamiltonian takes the form
\begin{eqnarray}
H_{\perp x} = H_{\perp x}^{G}[g_{ij},\pi^{ij}] + \tilde{H}_{\perp x}[\rho,\Phi;g_{ij}]~.\label{6 Super Hamiltonian split}
\end{eqnarray}

With this assumption in hand, it is possible to \emph{prove} \cite{Teitelboim thesis} that the metric appears in $\tilde{H}_{\perp x}$ only as a \emph{local} function of $g_{ij}$. That is, no derivatives of the metric are allowed, nor any other complicated functional dependencies; due to this fact, this was referred to as the  \emph{non-derivative} coupling assumption by Teitelboim.

\paragraph*{Modified Poisson brackets--- }
A particularly appealing aspect of this separation into gravitational and ``matter" pieces is that the Poisson bracket relations (\ref{6 PB 1})-(\ref{6 PB 3}) also split along similar lines. In fact, insert the decomposed generators $H_{Ax}$ from eqns.(\ref{6 Super momentum split}) and (\ref{6 Super Hamiltonian split}) into the Poisson bracket relations (\ref{6 PB 1})-(\ref{6 PB 3}). We find that the gravitational generators $H_{Ax}^{G} = (H_{\perp x}^{G},H_{ix}^{G})$ must satisfy a set of Poisson brackets
\begin{subequations}
\begin{eqnarray}
\left\{ H_{\bot x}^{G},H_{\bot x^{\prime }}^{G}\right\} &=&(g_{x}^{ij}H_{jx}^{G}+g_{x^{\prime
}}^{ij}H_{jx^{\prime }}^{G})\partial _{ix}\delta (x,x^{\prime })~,  \label{PB 1 G}
\\
\left\{ H_{ix}^{G},H_{\bot x^{\prime }}^{G}\right \} &=&H_{\bot x}^{G}\partial _{ix}\delta
(x,x^{\prime })~,  \label{PB 2 G} \\
\left\{ H_{ix}^{G},H_{jx^{\prime }}^{G}\right \} &=&H_{ix^{\prime }}^{G}\,\partial _{jx}\delta
(x,x^{\prime })+H_{jx}^{G}\,\partial _{ix}\delta (x,x^{\prime })~,  \label{PB 3 G}
\end{eqnarray}
\end{subequations}
which have closing relations \emph{identical} to those of (\ref{6 PB 1})-(\ref{6 PB 3}).

The ``matter" generators, which we collectively denote by $\tilde{H}_{Ax}$, satisfy a somewhat modified set of brackets
\begin{subequations}
\begin{eqnarray}
\left\{ \tilde{H}_{\bot x},\tilde{H}_{\bot x^{\prime }}\right\} &=&(g_{x}^{ij}\tilde{H}_{jx}+g_{x^{\prime
}}^{ij}\tilde{H}_{jx^{\prime }})\partial _{ix}\delta (x,x^{\prime })~,  \label{6 PB 1 matter}
\\
\left\{ H_{ix},\tilde{H}_{\bot x^{\prime }}\right \} &=&\tilde{H}_{\bot x}\partial _{ix}\delta
(x,x^{\prime })~,  \label{6 PB 2 matter} \\
\left\{ \tilde{H}_{ix},\tilde{H}_{jx^{\prime }}\right \} &=&\tilde{H}_{ix^{\prime }}\,\partial _{jx}\delta
(x,x^{\prime })+\tilde{H}_{jx}\,\partial _{ix}\delta (x,x^{\prime })~.  \label{6 PB 3 matter}
\end{eqnarray}
\end{subequations}
Here we call attention to the fact that eq.(\ref{6 PB 2 matter}) contains the total tangential generator $H_{ix}$, not just $\tilde{H}_{ix}$. This alteration occurs because the ``matter" super-Hamiltonian depends on the metric. That is, if we want to shift $\tilde{H}_{\perp x}$ along a given surface, we must shift the variables $(\rho,\Phi)$, as well as the metric $g_{ij}$; hence $H_{ix}$, not just $\tilde{H}_{ix}$, must appear.

Modulo the small distinction arising in eq.(\ref{6 PB 2 matter}), we see that the gravitational and ``matter" sectors decouple such that each piece forms an \emph{independent} representation of the ``algebra" of surface deformations. From a strategic point of view, the separation means that we can solve the Poisson bracket relations for geometry and ``matter" \emph{independently} of one another.

\paragraph*{The ``matter" super-Hamiltonian--- }
Our goal is to identify a family of ensemble super-Hamiltonians $\tilde{H}_{\perp x}[\rho,\Phi;g_{ij}]$ that are consistent with the Poisson brackets (\ref{6 PB 1 matter})-(\ref{6 PB 3 matter}). Let us briefly outline our approach. Thus far, we have completely determined the correct form of the ``matter" super-momentum $\tilde{H}_{ix}$ which is consistent with (\ref{6 PB 3 matter}). Moreover, the relation (\ref{6 PB 2 matter}) merely implies that $\tilde{H}_{\perp x}$ transforms as a scalar density under a spatial diffeomorphism. Thus we are left only to satisfy the first Poisson bracket (\ref{6 PB 1 matter}) for the unknown $\tilde{H}_{\perp x}$.

In addition to these considerations, however, the ED approach itself imposes additional constraints of a fundamental nature on the allowed $H_{\perp x}$, or more specifically $\tilde{H}_{\perp x}$. This is because, in ED, the introduction of a symplectic structure and its corresponding Hamiltonian formalism is not meant to replace the entropic updating methods that yield the LTFP equations, but to \emph{augment} them, appropriately. As a consequence, we demand, as a matter of principle, that the ``matter" super-Hamiltonian $\tilde{H}_{\perp x}$ be \emph{defined} so as to reproduce the LTFP equations of (\ref{FP equation}).

More explicitly, we require $\tilde{H}_{\perp x}$ to be such that its action on $\rho_{\sigma}$ generates the LTFP equations
\begin{eqnarray}
 \big\{ \rho_{\sigma}[\chi],H_{\perp x} \big\} =  \frac{\delta\rho_{\sigma}[\chi]}{\delta\xi^{\perp}_{x}}~, \label{FP equation H a}
\end{eqnarray}
which translates to
\begin{eqnarray}
\frac{\tilde{\delta}\tilde{H}_{\perp x}}{\tilde{\delta}\Phi_{\sigma}[\chi]}= -\frac{1}{g_{x}^{1/2}}\frac{\delta}{\delta\chi_{x}}\left (\rho_{\sigma}[\chi]\frac{\delta\Phi_{\sigma}[\chi]}{\delta\chi_{x}}    \right )~.\label{FP equation H b}
\end{eqnarray}
It is simple to check that the $\tilde{H}_{\perp x}$ that satisfies this condition is given by
\begin{eqnarray}
\tilde{H}_{\perp } = \int D\chi \rho_{\sigma} \frac{1}{2g_{x}^{1/2}}\left (  \frac{\delta\Phi_{\sigma}}{\delta\chi_{x}}\right )^{2} +F_{x}[\rho;g_{ij}]~.\label{e Hamiltonian a}
\end{eqnarray}
The first term in (\ref{e Hamiltonian a}) is \emph{fixed} by virtue of consistency with the LTFP equations, whereas $F_{x}[\rho;g_{ij}]$ is a yet undetermined ``constant" of integration, which may depend on $\rho$ as well as the metric. However, $F_{x}$ is not entirely arbitrary, its functional form is restricted by the Poisson bracket, eq.(\ref{6 PB 1 matter}).

Before proceeding, note that, up until this point, our discussion of path independence has been developed on a formal level, largely independent of ED itself. Such a formalism on its own, however, is necessarily devoid of many crucial physical ingredients. For instance, in ED we \emph{define} local time as a measure of the field fluctuations through (\ref{Fluctuations}), which leads to the LTFP equations. In principle, this has nothing to do with abstract parameters $\delta\xi^{A}_{x}$ introduced as part of local updating. It is only once we say that the entropic updating of ED must agree with the local time evolution generated by $\tilde{H}_{\perp x}$, made explicit in (\ref{FP b}), that the two notions coincide. In ED, the clock that measures the local proper time $\delta\xi^{\perp}_{x}$ is nothing but the field fluctuations themselves.

% It must be said that the decision to quantify the statistical state of the fields as a Bayesian probability need not imply that the truth

% The value of a quantity may be known for several reasons. It might be that such a quantity cannot be observed directly, such as a parameter in a model. Or it may be that repeated efforts to determine a value with precision, perhaps by measurement, turn out to be unsuccessful. The assignment of a probability is warranted in both cases.

Continuing with our task at hand, we seek a family of models that are consistent with eq.(\ref{6 PB 1 matter}). This is accomplished for suitable choices of $F_{x}$. We pursue this in manner similar to \cite{Ipek et al 2019}. Begin by rewriting $\tilde{H}_{\perp x}$ as
\begin{eqnarray}
\tilde{H}_{\perp x} = \tilde{H}_{\perp x}^{0}+ F_{x}~, \label{6 e-H}
\end{eqnarray}
where we have introduced
\begin{eqnarray}
\tilde{H}_{\perp x}^{0} = \int D\chi \,\rho_{\sigma}\left ( \frac{1}{%
2g_{x}^{1/2}}\left( \frac{\delta \Phi_{\sigma} }{\delta \chi _{x}}\right) ^{2}+\frac{g^{1/2}_{x}}{2}g^{ij}\partial_{ix}\chi_{x}\partial_{jx}\chi_{x}\right ) ~.  \label{6 e-H b}
\end{eqnarray}%
This amounts simply to a redefinition of the arbitrary $F_{x}$ in (\ref{e Hamiltonian a}). An advantage of this definition, however, is that the newly defined $\tilde{H}_{\perp x}^{0}$ automatically satisfies
\begin{eqnarray}
\left\{ \tilde{H}^{0}_{\bot x},\tilde{H}^{0}_{\bot x^{\prime }}\right\} =(g_{x}^{ij}\tilde{H}_{jx}+g_{x^{\prime
}}^{ij}\tilde{H}_{jx^{\prime }})\partial _{ix}\delta (x,x^{\prime })~.  \label{6 PB 1 matter Ho}
\end{eqnarray}
Finding a suitable $F_{x}$ is therefore accomplished by satisfying\footnote{Since $F_{x}[\rho; g_{ij}]$ is independent of $\Phi$, we have that $\left\{ F_{x},F_{x^{\prime}}\right\} = 0 $, identically, from which eq.(\ref{PB 1 matter necessary}) follows.}
\begin{eqnarray}
\left\{ \tilde{H}^{0}_{\bot x},F_{x^{\prime}}\right\} = \left\{ \tilde{H}^{0}_{\bot x^{\prime}},F_{x}\right\}~.\label{PB 1 matter necessary}
\end{eqnarray}
Clearly a necessary condition for an acceptable $F_{x}$ is that the Poisson bracket $\left\{ \tilde{H}^{0}_{\bot x},F_{x^{\prime}}\right\} $ must be symmetric upon exchange of $x$ and $x^{\prime}$. Since $\tilde{H}^{0}_{\perp x}$ must itself reproduce the LTFP equations, the condition (\ref{PB 1 matter necessary}) translates to \cite{Ipek et al 2017}\cite{Ipek et al 2019}
\begin{eqnarray}
\frac{1}{g_{x}^{1/2}}\frac{\delta}{\delta\chi_{x}}\left (\rho_{\sigma} \frac{\delta}{\delta\chi_{x}} \frac{\tilde{\delta}F_{x^{\prime}}}{\tilde{\delta}\rho_{\sigma}}   \right ) = \frac{1}{g_{x^{\prime}}^{1/2}}\frac{\delta}{\delta\chi_{x^{\prime}}}\left (\rho_{\sigma} \frac{\delta}{\delta\chi_{x^{\prime}}} \frac{\tilde{\delta}F_{x}}{\tilde{\delta}\rho_{\sigma}}   \right )~,\label{PB 1 matter necessary b}
\end{eqnarray}
which is an equation linear in $F_{x}$.

A complete description of solutions to eq.(\ref{PB 1 matter necessary b}) lies outside the scope of the current work. However, a restricted family of solutions, which are nonetheless of physical interest, can be found for $F_{x}$'s of the form
\begin{eqnarray}
F_{x}[\rho] = \int D\chi f_{x}\left(\rho,\frac{\delta\rho}{\delta\chi_{x}}; g_{ijx}\right)~,\label{6 Local potentials}
\end{eqnarray}
where $f_{x}$ is a \emph{function}, not functional, of its arguments. For such a special type of $F_{x}$ one can check by substitution into (\ref{PB 1 matter necessary b}) that
\begin{eqnarray}
f_{x} \sim g_{x}^{1/2}\rho \chi_{x}^{n}\quad \text{(integer n)}\quad \text{and}\quad f_{x}\sim \frac{\rho}{g_{x}^{1/2}}\left (\frac{\delta\log\rho}{\delta\chi_{x}}\right )^{2}\notag
\end{eqnarray}
are acceptable solutions. Since eq.(\ref{PB 1 matter necessary b}) is \emph{linear} in $F_{x}$, solutions can be superposed so that a suitable family of $\tilde{H}_{\perp x}$'s is given by
\begin{eqnarray}
\tilde{H}_{\perp x} &=& \int D\chi \, \rho_{\sigma} \left ( \frac{1}{%
2g_{x}^{1/2}}\left( \frac{\delta \Phi_{\sigma} }{\delta \chi _{x}}\right) ^{2}+\frac{g^{1/2}_{x}}{2}g^{ij}\partial_{ix}\chi_{x}\partial_{jx}\chi_{x}\right.  \notag\\
&+&\left. g_{x}^{1/2}V_{x}(\chi_{x})+ \frac{\lambda}{g_{x}^{1/2}}\left (\frac{\delta\log\rho_{\sigma}}{\delta\chi_{x}}\right )^{2}\right )~,\label{e Hamiltonian final}
\end{eqnarray}
where $V_{x} = \sum_{n}\lambda_{n}\chi_{x}^{n}$ is a function that is polynomial in $\chi_{x}$.

As discussed in \cite{Ipek et al 2019}, the last term can be interpreted as the ``local quantum potential". To see this we recall that in flat space-time the quantum potential is given by \cite{Ipek Caticha 2014}
\begin{eqnarray}
Q =\int d^{3}x \int D\chi \rho \, \lambda \left (\frac{\delta\log\rho}{\delta\chi_{x}}\right )^{2}~.\label{Quantum Potential Flat}
\end{eqnarray}
The transition to a curved space-time is made by making the substitutions
\begin{eqnarray}
d^{3}x \to g_{x}^{1/2}d^{3}x\quad\text{and}\quad \frac{\delta}{\delta\chi_{x}}\to\frac{1}{g_{x}^{1/2}}\frac{\delta}{\delta\chi_{x}}~,\notag
\end{eqnarray}
yielding the result
\begin{eqnarray}
Q_{\sigma} = \int d^{3}x \int D\chi \rho  \frac{\lambda}{g_{x}^{1/2}} \left (\frac{\delta\log\rho}{\delta\chi_{x}}\right )^{2}~.\notag
\end{eqnarray}
The term inside the spatial integral is exactly what appears in (\ref{e Hamiltonian final}), which justifies the name. The contribution of the local quantum potential to the energy is such that those states that are more smoothly spread out in configuration space tend to have lower energy. The corresponding coupling constant $\lambda > 0$ controls the relative importance of the quantum potential; the case $\lambda < 0$ is excluded because it leads to instabilities.

\paragraph*{The gravitational super-Hamiltonian--- }
We now proceed to determining the last remaining element of our scheme, the gravitational super-Hamiltonian $H_{\perp x}^{G}$. Fortunately, under the assumption of non-derivative coupling mentioned above, it is possible to completely separate the Poisson bracket relations (\ref{6 PB 1})-(\ref{6 PB 3}) into pieces that are purely gravitational and those that are pure ``matter". Consequently, to determine $H_{\perp x}^{G}$ it suffices merely to solve the Poisson bracket relations (\ref{PB 1 G})-(\ref{PB 3 G}), which involve only the gravitational variables $g_{ij}$ and $\pi^{ij}$.

Such a task, however, is mathematically equivalent to determining the generators of \emph{pure} geometrodynamics, in which the coupling to ``matter" is absent. But it is precisely this latter challenge which was, in fact, addressed by the efforts of HKT in \cite{Hojman Kuchar Teitelboim 1976} --- they proposed a solution to exactly those brackets that appear in (\ref{PB 1 G})-(\ref{PB 3 G}). Their main result was an important one: the only time-reversible solution to equations (\ref{PB 1 G})-(\ref{PB 3 G}) is nothing less than Einstein's GR in vacuum. We briefly review their argument here, and adapt their solution to our current work in ED.

To begin, since $g_{ij}$ is the intrinsic metric of the surface, its behavior under a purely normal deformation is known \cite{Gourgoulhon 2007} to be
\begin{eqnarray}
\delta g_{ij}(x) = \pounds_{\delta \xi^{\perp}}g_{ijx} = -2 K_{ijx}~,\label{Normal deformation metric a}
\end{eqnarray}
where $K_{ij}(x) = K_{ijx}$ is a symmetric $(0,2)$ tensor that is called the \emph{extrinsic} curvature. Note that identifying $K_{ijx}$ with the response of $g_{ijx}$ under a normal deformation is a geometric requirement, not a dynamical one; without this, we could not interpret $g_{ij}$ as residing on a space-like cut of space-time. That being said, $K_{ijx}$ is at this juncture an undetermined functional of the canonical variables, and more to the point, we do not assume at the moment any simple relationship with the momenta $\pi^{ij}_{x}$ --- this must be derived.

Alternatively, the deformation in (\ref{Normal deformation metric a}) must also be attainable using the normal generator $H_{\perp x}^{G}$,
\begin{eqnarray}
\delta g_{ijx} = \int dx^{\prime}\left\{ g_{ijx},H_{\perp x^{\prime}}^{G} \right \}\delta\xi^{\perp}_{x^{\prime}}~.\notag
\end{eqnarray}
More explicitly $H_{\perp x}^{G}$ should satisfy
\begin{eqnarray}
\frac{\delta H_{\perp x^{\prime}}^{G}}{\delta\pi^{ij}_{x}} =-2 K_{ijx}\delta(x,x^{\prime})~. \label{6 Normal deformation metric}
\end{eqnarray}
The appearance of the Dirac delta in eq.(\ref{6 Normal deformation metric}) implies that $H_{\perp x}^{G}$ is \emph{local} in $\pi^{ij}_{x}$. Therefore $H_{\perp x}^{G}$ is a \emph{function}, not a functional, of $\pi^{ij}_{x}$.

Such a simplification turns out to be important: once $H_{\perp x}^{G}$ is a function of $\pi^{ij}$, it is possible to produce an \emph{ansatz} for $H_{\perp x}^{G}$ in terms of powers of $\pi^{ij}$. HKT then supplement this with an additional simplifying assumption: that geometrodynamics be \emph{time-reversible}. This has the advantage of removing all terms in $H_{\perp x}^{G}$ that have odd powers of $\pi^{ij}_{x}$.

%The appearance of the Dirac delta in eq.(\ref{6 Normal deformation metric}) is valuable from a strategic point of view, as it requires that $H_{\perp x}^{G}$ be \emph{local} in $\pi^{ij}_{x}$. We can therefore consider $H_{\perp x}^{G}$ to be a \emph{function}, not functional, of $\pi^{ij}_{x}$.

Incorporating these ingredients, it is then possible to consider $H_{\perp x}^{G}$'s of the form
\begin{eqnarray}
H_{\perp x}^{G} = \sum_{n = 0}^{\infty}G^{(2n)}_{i_{1}j_{1}i_{2}j_{2}\cdots i_{2n}j_{2n}x}\pi^{i_{1}j_{1}}_{x}\pi^{i_{2}j_{2}}_{x}\cdots\pi^{i_{2n}j_{2n}}_{x}~,\label{6 Super hamiltonian grav ansatz}
\end{eqnarray}
where the coefficients $G^{(2n)}_{ij\cdots}$ are functionals of the metric $g_{ij}$, but depend on the point $x$. Furthermore, since $H_{\perp x}^{G} $ is scalar density and $\pi^{ij}$ a tensor density, $G_{ij\cdots}^{(2n)}$ must transform as a tensor density of weight $1-2n$. Since $\pi^{ij}$ is a symmetric tensor we expect $G^{(2n)}_{ij\cdots}$ to be symmetric under exchange of indices $i_{a}j_{a}\leftrightarrow j_{a}i_{a}$ for any pair $a$; also, $G^{(2n)}_{ij\cdots}$ should be symmetric on interchange of any pair $i_{a}j_{a}\leftrightarrow i_{b}j_{b}$ as this just corresponds to exchanging the $\pi$'s.

Naturally, one narrows the allowable $H_{\perp x}^{G}$ by inserting the ansatz (\ref{6 Super hamiltonian grav ansatz}) into the Poisson bracket (\ref{PB 1 G}). Without delving too far into the details (which can be found in \cite{Hojman Kuchar Teitelboim 1976}) we quote their solution. Only the $n = 0, 1$ terms survive, giving
\begin{eqnarray}
H_{\perp x}^{G} = \kappa \, G_{ijkl}\pi^{ij}\pi^{kl} - \frac{g^{1/2}}{2\kappa}\left(R-\Lambda\right)~,
 \label{6 Super hamiltonian grav cosmo}
\end{eqnarray}
where we have introduce the super metric
\begin{eqnarray}
G_{ijkl} =  \frac{1}{g_{x}^{1/2}}\left(g_{ik}g_{jl}+g_{il}g_{jk}-g_{ij}g_{kl}\right)~,
\end{eqnarray}
and where $R$ is the Ricci scalar for the metric $g_{ij}$.

The constant $\kappa$ is a coefficient that, eventually, determines the coupling to ``matter". We follow standard convention in identifying it as $\kappa = 8\pi G$, where $G$ is Newton's constant. The other parameter $\Lambda$, of course, is the cosmological constant. For simplicity, going forward we set $\Lambda = 0$.

Putting everything together, we have that $H_{\perp x}^{G}$ takes the form
\begin{eqnarray}
H_{\perp x}^{G} = \frac{\kappa}{g_{x}^{1/2}}\left (2\pi^{ij}\pi_{ij}-\pi^{2}\right ) - \frac{g^{1/2}}{2\kappa}R~,\label{6 Super hamiltonian grav}
\end{eqnarray}
where $\pi = \pi^{ij}g_{ij} = \text{Tr}(\pi^{ij}) $. This is exactly the standard gravitational super-Hamiltonian obtained by Dirac \cite{Dirac 1958} and ADM \cite{ADM 1960} by starting from the Einstein-Hilbert Lagrangian.

\paragraph*{Total super-Hamiltonian--- }
Putting together the ingredients of this section, the total super-Hamiltonian is thus
\begin{eqnarray}
H_{\perp x} = H^{G}_{\perp x}+\tilde{H}_{\perp x} ~,\label{6 Super-Hamiltonian Total}
\end{eqnarray}
where $H_{\perp x}^{G}$ is given above by eq.(\ref{6 Super hamiltonian grav}) and where a suitable family of $\tilde{H}_{\perp x}$'s have been identified in eq.(\ref{6 e-H}). With this, the super-Hamiltonian constraint is then just
\begin{eqnarray}
H_{\perp x} = H^{G}_{\perp x}+\tilde{H}_{\perp x} \approx 0~.\label{6 Super-Hamiltonian constraint}
\end{eqnarray}

\section{The dynamical equations}\label{ED_DyEqns}
%In the previous section we have identified a representation of the relations eqns.(\ref{6 PB 1})-(\ref{6 Hamiltonian constraints}) in terms of the canonical variables $(\rho,\Phi;g_{ij},\pi^{ij})$. We now investigate the dynamical equations induced by these generators, the $H_{Ax}$.

As argued by HKT, although the role of space-time was crucial to the developing the equations (\ref{6 PB 1})-(\ref{6 Hamiltonian constraints}), one notices that all signs of the enveloping space-time have dropped out in the closing relations (\ref{6 PB 1})-(\ref{6 PB 3}).\footnote{For instance, they depend on the surface metric $g_{ij}$, but not on its extrinsic curvature $K_{ij}$.} Thus in geometrodynamics we can dispense with the notion of an \emph{a priori} given space-time and instead consider the three-dimensional Riemannian manifold $\sigma_{t}$ as primary.

The idea is a simple one. The canonical variables are evolved with the generators $H_{Ax}$, satisfying eqns.(\ref{6 PB 1})-(\ref{6 Hamiltonian constraints}). Such an evolution will cause both the ``matter" and geometry, to change. We might then give this manifold with updated intrinsic geometry a new name, $\sigma_{t'}$. Repeating this procedure results in an evolution of the dynamical variables, and consequently, what one might view as a succession of manifolds $\{\sigma_{t}\}$, parameterized by the label $t$. Thus this iterative process constructs a space-time, step by step. We now investigate the dynamical equations that result from this procedure.

\subsection*{Some formalism}
Consider the evolution of an arbitrary functional $T_{t}$ of the dynamical variables defined on $\sigma_{t}$, 
\begin{eqnarray}
\delta T_{t} = T_{t+dt} - T_{t} = \int dx \left \{T_{t}, H_{Ax} \right \}\delta\xi^{A}_{x} =  \int dx \, N^{A}_{xt} \left \{T_{t}, H_{Ax} \right \}dt~,\label{6 Evolution lapse shift}
\end{eqnarray}
where we have introduced four arbitrary functions
\begin{eqnarray}
N_{xt} = \frac{\delta\xi^{\perp}_{x}}{dt}\quad\text{and}\quad N^{i}_{xt} = \frac{\delta\xi^{i}_{x}}{dt}~,\label{6 Def Lapse Shift}
\end{eqnarray}
which are the \emph{lapse} and vector \emph{shift}, respectively. Just as the evolution parameters $\delta\xi^{A}_{x}$ were completely arbitrary, these functions can be freely specified, and amount \emph{eventually} to picking a particular foliation of space-time.\footnote{At the moment, the lapse and shift have no definite geometrical meaning. But as is well known, we can eventually identify the lapse $N$ and shift $N^{i}$ as being related to components of the space-time metric $^{4}g_{\mu\nu} \equiv \gamma_{\mu\nu}$. (We changed the symbol temporarily to avoid confusion.) More specifically, we would have \cite{Hojman Kuchar Teitelboim 1976}
\begin{eqnarray}
N = \left(- \gamma^{00}\right)^{-1/2}\quad\text{and}\quad N_{i} = \gamma_{0i}~.\notag
\end{eqnarray}
}

As in the Dirac approach to geometrodynamics, the $N^{A}_{xt} = (N_{xt}, N^{i}_{xt})$ are \emph{not} functionals of the canonical variables, therefore we can rewrite eq.(\ref{6 Evolution lapse shift}) as
\begin{eqnarray}
\delta T_{t}  = \left \{T_{t}, H[N,N^{i}] \right \}dt~,\label{6 Evolution smeared}
\end{eqnarray}
where we have introduced the notion of an ``integrated", or ``smeared" Hamiltonian
\begin{eqnarray}
H[N,N^{i}]=\int dx\,\left( N_{xt}H_{\bot x}+N_{xt}^{i}H_{ix}\right) \label{6 Smeared Hamiltonian}
\end{eqnarray}%
that, for given $N^{A}_{xt}$, generates an evolution parameterized by $t$.\footnote{Had the $N^{A}$ been functionals of the canonical variables then the Poisson bracket in eq.(\ref{6 Evolution smeared}) would have generated extra terms from their action on $N^{A}$ and thus eq.(\ref{6 Evolution smeared}) would not have been equivalent to eq.(\ref{6 Evolution lapse shift}).} This \emph{global} $H[N,N^{i}]$ (note the spatial integral) conforms more naturally to our typical notions of a Hamiltonian. Thus, the derivative with respect to the parameter $t$ is
\begin{eqnarray}
\partial_{t} T_{t} \equiv \frac{\delta T_{t}}{dt} = \left \{ T_{t}, H[N,N^{i}]    \right \}~,\label{6 Def time derivative}
\end{eqnarray}
or more explicitly by
\begin{eqnarray}
\partial_{t} T_{t} = \int dx\,\left( N_{xt}\left \{T_{t},H_{\perp x}\right \}+N_{xt}^{i}\left \{T_{t},H_{ix}\right \}\right)~,\label{6 Def time derivative b}
\end{eqnarray}
and more succinctly
\begin{eqnarray}
\partial_{t} T_{t} = \pounds_{m} T_{t} + \pounds_{N^{i}} T_{t}~, \label{6 Def time derivative c}
\end{eqnarray}
where the vector $m^{\mu}_{x} = N_{xt} n^{\mu}_{x}$ is the so-called evolution vector.

\subsection*{The evolution of the ``matter" sector}
The goal is to determine the evolution of the probability distribution $\rho_{t}[\chi]$ and phase functional $\Phi_{t}[\chi]$, given an initial state $(\rho_{t}, \Phi_{t}; g_{ij\, t}, \pi^{ij}_{t})$ that satisfies the initial value constraints (\ref{6 Hamiltonian constraints}).
\paragraph*{Dynamical equations for the probability and phase--- }
Given the basic dynamical law, eq.(\ref{6 Def time derivative}), the time evolution of the variables $\rho_{t}$ and $\Phi_{t}$ are given by
\begin{eqnarray}
\partial_{t}\rho_{t} = \left \{\rho_{t}, H   \right \}\quad\text{and}\quad \partial_{t}\Phi_{t} = \left \{\rho_{t}, H   \right \}~,
\end{eqnarray}
where $H$ is the smeared Hamiltonian given above. Thus, in accordance with eq.(\ref{6 Def time derivative b}), we have the result
\begin{subequations}
\begin{eqnarray}
\partial_{t}\rho_{t}  &=& \int dx \, \left ( N_{xt}\left \{\rho_{t}, H_{\perp x}\right \}+ N^{i}_{xt}\left \{\rho_{t}[\chi], H_{i x}\right \} \right ) \\ 
\partial_{t}\Phi_{t}  &=& \int dx \, \left ( N_{xt}\left \{\Phi_{t}, H_{\perp x}\right \}+ N^{i}_{xt}\left \{\Phi_{t}, H_{i x}\right \} \right )~. \label{6 Evolution Phi}
\end{eqnarray}
\end{subequations}

Using the family of $H_{\perp x}$'s that we identified in eq.(\ref{6 Super-Hamiltonian Total}) and the super-momentum $H_{ix}$ in eq.(\ref{6 Super momentum}) we can compute all of the necessary Poisson brackets. From the super momentum in eq.(\ref{6 Super momentum}) we can compute the tangential pieces
\begin{subequations}
\begin{eqnarray}
\left \{\rho_{t}[\chi], H_{i x}\right \} &=& \frac{\tilde{\delta}\tilde{H}_{ix}}{\tilde{\delta}\tilde{\Phi_{t}}} = \frac{\delta\rho_{t}}{\delta\chi_{x}}\partial_{ix}\chi_{x}~,\\
\left \{\Phi_{t}[\chi], H_{i x}\right \} &=& -\frac{\tilde{\delta}\tilde{H}_{ix}}{\tilde{\delta}\tilde{\rho_{t}}} = \frac{\delta\Phi_{t}}{\delta\chi_{x}}\partial_{ix}\chi_{x}~.\label{6 Tangential PB}
\end{eqnarray}
\end{subequations}
Of course, the entire family of $\tilde{H}_{\perp x}$'s were designed to reproduce the LTFP equations, thus by construction we have
\begin{eqnarray}
\left \{\rho_{t}[\chi], H_{\perp x}\right \} = \frac{\tilde{\delta}\tilde{H}_{\perp x}}{\tilde{\delta}\tilde{\Phi_{t}}} = -\frac{1}{g_{x}^{1/2}}\frac{\delta}{\delta\chi_{x}}\left ( \rho_{t}\frac{\delta\Phi_{t}}{\delta\chi_{x}}  \right )~,\label{FP equation final}
\end{eqnarray}
in agreement with eq.(\ref{FP equation}). The remaining Poisson bracket determines the local time evolution of the phase functional $\Phi_{t}$ and is given by
\begin{align}
\left \{\Phi_{t}[\chi], H_{\perp x}\right \} = - \frac{\tilde{\delta}\tilde{H}_{\perp x}}{\tilde{\delta}\tilde{\rho_{t}}}  = \frac{1}{2 g_{x}^{1/2}}\left(\frac{\delta\Phi_{t}[\chi]}{\delta\chi_{x}}\right)^{2}+\frac{g_{x}^{1/2}}{2}g^{ij}\partial_{i}\chi_{x}\partial_{j}\chi_{x}+ \frac{\tilde{\delta}F_{x}[\rho]}{\tilde{\delta}\rho_{t}[\chi]}~,\label{6 LTHJ equations}
\end{align}
where $F_{x}$ is of the form specified by eq.(\ref{6 e-H}).
\paragraph*{The local time Hamilton-Jacobi equations--- }
To interpret the local equations (\ref{6 LTHJ equations}), we write the full time evolution for the phase functional by inserting eqns.(\ref{6 LTHJ equations}) and (\ref{6 Tangential PB}) into eq.(\ref{6 Evolution Phi}), yielding
\begin{align}
-\partial_{t}\Phi_{t} =& \int dx \, N_{xt} \left (\frac{1}{2 g_{x}^{1/2}}\left(\frac{\delta\Phi_{t}}{\delta\chi_{x}}\right)^{2} +\frac{g_{x}^{1/2}}{2}g^{ij}\partial_{i}\chi_{x}\partial_{j}\chi_{x} + \frac{\tilde{\delta}F_{x}}{\tilde{\delta}\rho_{t}}\right )\notag\\
&+\int dx \, N^{i}_{xt}\frac{\delta\Phi_{t}}{\delta\chi_{x}}\partial_{ix}\chi_{x}~.
\end{align}
To bring this equation into a more familiar form we consider the special case of flat space-time by setting the metric to be a Kroenecker delta $g_{ij} = \delta_{ij}$, so that  $g_{x}^{1/2} = 1$, and we let  $N = 1$, $N^{i} = 0$. Moreover, for simplicity we also make the assignment $F_{x} = 0$. This results in a time evolution for $\Phi_{t}$ that has the form
\begin{eqnarray}
-\partial_{t}\Phi_{t} = \int dx \left (\frac{1}{2}\left(\frac{\delta\Phi_{t}}{\delta\chi_{x}}\right)^{2} +\frac{1}{2}\delta^{ij}\partial_{i}\chi_{x}\partial_{j}\chi_{x} \right )~,\notag
\end{eqnarray}
which is exactly the classical Hamilton-Jacobi equation for a massless Klein-Gordon field in flat space-time. Thus, in analogy with the LTFP equations, we refer to eq.(\ref{6 LTHJ equations}) as the \emph{local time Hamilton-Jacobi} (LTHJ) equations, as there is one equation of the Hamilton-Jacobi type for every spatial point.

The LTFP and LTHJ equations, with the tangential equations (\ref{6 Tangential PB}), and the evolution equations (\ref{6 Evolution Phi}), give us the ability to evolve an appropriately chosen initial state $(\rho_{t},\Phi_{t})$. In general, this is a coupled non-linear evolution driven by a dependence on the metric $g_{ij}$. To a large extent, this completes our discussion of how the epistemic variables evolve. In a subsequent section, however, we discuss in some detail the dynamics of a specific class of models, those that involve the local quantum potential.

\subsection*{The evolution of the geometrical variables}
We now review the content of  the well-known Einstein's equations written within the canonical language. (A good review of these equations is given, for example, by ADM \cite{ADM 2008}.)
\paragraph*{Evolution of metric--- }
The goal is to determine the evolution of the geometrical variables $(g_{ij},\pi^{ij})$. Beginning with the metric $g_{ij}$ defined on some initial three-space $\sigma_{t}$, we wish to determine how it evolves in response to the generators $H_{Ax}$. Applying  eq.(\ref{6 Def time derivative b}) for the time derivative, we have that
\begin{eqnarray}
\partial_{t} g_{ijx} = \int dx^{\prime} \left\{g_{ijx} , H_{Ax'} \right \}N^{A}_{x^{\prime}t} =  \int dx^{\prime} \frac{\delta H_{Ax'}}{\delta\pi^{ij}_{x}}N^{A}_{x^{\prime}t}  ~.\label{6 Metric time deriv a}
\end{eqnarray}
To compute this, recall that the tangential piece is known from equations (\ref{6 Tangential deformation metric}) and (\ref{6 Tangential deformation metric cov}). This gives us
\begin{eqnarray}
\pounds_{N^{i}} g_{ijx} &=  \int dx^{\prime}  \frac{\delta H_{ix'}}{\delta\pi^{ij}_{x}}N^{i}_{x^{\prime}t} = \nabla_{i}N_{jxt}+\nabla_{j}N_{ixt}~, \label{6 Lie derivative shift  metric}
\end{eqnarray}
where $\pounds_{N^{i}} g_{ijx}$ is just the Lie derivative along the vector field defined by the shift $N^{i}$, and where $N_{ixt} = g_{ijx}N^{j}_{xt}$.

To obtain the remaining piece, first differentiate the $H_{\perp x}$ given in eq.(\ref{6 Super hamiltonian grav})
\begin{eqnarray}
 \frac{\delta H_{\perp x^{\prime}}}{\delta\pi^{ij}_{x}} = \frac{4\kappa}{g_{x}^{1/2}}\left(\pi_{ijx} - \frac{1}{2}\pi_{x} g_{ijx}\right)\delta(x,x^{\prime})~,\notag%\label{6 Normal deformation metric b}
\end{eqnarray}
where $\pi_{ij}$ is the conjugate momentum with its indices lowered, and $\pi = g_{ij}\pi^{ij} = \text{Tr}(\pi^{ij})$ is the trace of the gravitational momentum. Putting these terms together, eq.(\ref{6 Metric time deriv a}) reads as
\begin{subequations}
\begin{eqnarray}
\partial_{t}g_{ijx} &=& \pounds_{m}g_{ijx} + \pounds_{N^{i}}g_{ijx} \label{6 Metric time deriv b}\\
 \pounds_{m}g_{ijx} &=&  \frac{2\kappa}{g_{x}^{1/2}}\left(2\pi_{ijx} - \pi_{x} g_{ijx}\right)N_{xt} \\
  \pounds_{N^{i}}g_{ijx} &=& \nabla_{i}N_{jxt}+\nabla_{j}N_{ixt}~.
\end{eqnarray}
\end{subequations}
This gives us the evolution of the metric with foliation parameter $t$.

We can also now identify the extrinsic curvature tensor
\begin{eqnarray}
K_{ij} = \frac{\kappa}{g^{1/2}_{x}}\left(\pi_{x} g_{ijx} - 2\pi_{ijx}\right) ~.
\end{eqnarray}
Inverting this relationship we get $\pi^{ij}$ in terms of $K^{ij}$, which yields
\begin{eqnarray}
\pi^{ij}_{x} = \frac{g^{1/2}_{x}}{2\kappa}\left (Kg^{ij}_{x} - K^{ij}_{x} \right )~. \label{6 Extrinsic curvature pi b}
\end{eqnarray}
where $K = K_{ij}g^{ij} = \text{Tr}(K_{ij})$ is the trace of $K_{ij}$. This is of some interest if one wishes to compare the canonical formulation to the standard so-called ``Lagrangian" approach (see e.g., \cite{Carroll 2004}).

\paragraph*{Evolution of conjugate momentum--- }
To this point, the dynamics of the geometry has not differed from a purely classical geometrodynamics, such as that of HKT. This is because the ``matter" super-Hamiltonian and super-momentum were, by definition, completely independent of $\pi^{ij}$ (due, of course, to the non-derivative coupling assumption) therefore the evolution of $g_{ij}$ did not receive contributions from the ``matter" sector. This changes when we consider the dynamics of the conjugate momentum $\pi^{ij}$.

The evolution of $\pi^{ij}$ is determined \emph{via} the equation
\begin{eqnarray}
\partial_{t} \pi_{x}^{ij} = \int dx^{\prime} \left\{\pi^{ij}_{x} , H_{Ax'} \right \}N^{A}_{x^{\prime}t} =  - \int dx^{\prime} \frac{\delta H_{Ax'}}{\delta g_{ijx}}N^{A}_{x^{\prime}t}  ~.\label{6 Pi time deriv a}
\end{eqnarray}
Recalling now that both the gravitational and ``matter" super-Hamiltonians $H_{\perp x} = H_{\perp x}^{G}[g_{ij},\pi^{ij}] + \tilde{H}_{\perp x}[\rho,\Phi;g_{ij}]$ depend explicitly on the metric, but that $\tilde{H}_{ix}$ does not, this expression slightly simplifies to
\begin{eqnarray}
\partial_{t}\pi^{ij}_{x}= -\int dx^{\prime}\left(N_{x^{\prime}t}\frac{\delta H^{G}_{\perp x^{\prime}}}{\delta g_{ijx}}+ N_{x^{\prime}t}\frac{\delta \tilde{H}_{\perp x^{\prime}}}{\delta g_{ijx}} +N^{i}_{x^{\prime}t}\frac{\delta H_{ix^{\prime}}^{G}}{\delta g_{ijx}}\right)~,
\end{eqnarray}
where we have have separated the contributions from the gravitational and ``matter" sectors. In the notation of eq.(\ref{6 Def time derivative c}) we write this as
\begin{eqnarray}
\partial_{t}\pi^{ij}_{x} =  \pounds_{m}\pi^{ij}_{x} +\pounds_{N^{i}} \pi^{ij}_{x}~,
\end{eqnarray}
where
\begin{eqnarray}
\pounds_{m}\pi^{ij}_{x} = \pounds_{m}^{G}\pi^{ij}_{x}+ \pounds_{m}^{M}\pi^{ij}_{x}~,
\end{eqnarray}
and
\begin{subequations}
\begin{eqnarray}
\pounds_{m}^{G}\pi^{ij}_{x} &=& -\int dx^{\prime}\left(N_{x^{\prime}t}\frac{\delta H^{G}_{\perp x^{\prime}}}{\delta g_{ijx}}\right)~,  \label{6 Pi Lie normal G} \\ 
\pounds_{m}^{M}\pi^{ij}_{x} &=& -\int dx^{\prime}\left(N_{x^{\prime}t}\frac{\delta \tilde{H}_{\perp x^{\prime}}}{\delta g_{ijx}}\right)~, \label{6 Pi Lie normal M} \\
\pounds_{N^{i}}\pi^{ij}_{x} &=& -\int dx^{\prime}\left(N^{i}_{x^{\prime}t}\frac{\delta H^{G}_{i x^{\prime}}}{\delta g_{ijx}}\right)~.\label{6 Pi Lie tangential G}
\end{eqnarray}
\end{subequations}

From (\ref{6 Tangential deformation momentum cov}) the last term is fairly easy to determine,
\begin{eqnarray}
\pounds_{N^{i}}\pi^{ij}_{x} =  \nabla_{kx}\left(\pi^{ij}_{x}N^{k}_{x}\right) - \pi^{ik}_{x}\nabla_{kx}N^{j}_{x} - \pi^{kj}_{x}\nabla_{kx}N^{i}_{x}
~.\label{6 Pi Lie tangential G b}
\end{eqnarray}

The calculation of the other two terms is much more involved. Fortunately the expression for $\pounds_{m}^{G}\pi^{ij}$ is already well known \cite{Gourgoulhon 2007}\cite{ADM 2008} and we merely quote the result,
\begin{align}
\pounds_{m}^{G}\pi^{ij}_{x} &=-\frac{g^{1/2}}{2\kappa}\left(R^{ij}_{x}-\frac{1}{2}g^{ij}_{x}R_{x}\right )N_{xt}+ \frac{\kappa}{g^{1/2}_{x}} \, g^{ij}_{x}\left(\pi^{kl}_{x}\pi_{klx}-\frac{1}{2}\pi^{2}_{x}\right)N_{xt}\notag\\
 &-\frac{4\kappa}{g^{1/2}_{x}}\left(\pi^{ik}_{x}\pi^{j}_{kx}-\frac{1}{2}\pi_{x}\pi^{ij}_{x}\right)N_{xt}+\frac{g^{1/2}_{x}}{2\kappa}\left(\nabla^{i}_{x}\nabla^{j}_{x}N_{xt} - g^{ij}_{x}\nabla^{k}_{x}\nabla_{kx}N_{xt}\right)~.\label{Lie derivative normal pi G}
\end{align}

To calculate the remaining piece $\pounds_{m}^{M}\pi^{ij}_{x}$ we first recall that having assumed a \emph{non-derivative} coupling of gravity to matter, the metric $g_{ij}$ appears in $\tilde{H}_{\perp x}$ as an undifferentiated function (not a functional) and without any derivatives \cite{Teitelboim thesis}, which implies that
\begin{eqnarray}
\frac{\delta \tilde{H}_{\perp x^{\prime}}}{\delta g_{ijx}} =\frac{\partial \tilde{H}_{\perp x}}{\partial g_{ijx}} \,  \delta(x,x') ~,\notag
\end{eqnarray}
obtaining
\begin{eqnarray}
\pounds_{m}^{M}\pi^{ij}_{x} = - \frac{\partial \tilde{H}_{\perp x}}{\partial g_{ijx}} N_{xt}~.\notag
\end{eqnarray}
Using eq.(\ref{6 e-H}), for an arbitrary choice of $F_{x}$, but one that still satisfies eq.(\ref{6 PB 1 matter}), the ``matter" source has the form
\begin{eqnarray}
\pounds_{m}^{M}\pi^{ij}_{x}  = N_{xt} \left (  \frac{\partial \tilde{H}^{0}_{\perp x}}{\partial g_{ijx}} + \frac{\partial F_{x}}{\partial g_{ijx}}  \right )~,\label{6 Matter source}
\end{eqnarray}
with
\begin{align}
 \frac{\partial \tilde{H}^{0}_{\perp x}}{\partial g_{ijx}} =& -\frac{1}{2}\int D\chi \rho \left (\frac{1}{g_{x}^{1/2}}\left (\frac{\delta\Phi}{\delta\chi_{x}}\right )^{2}  -g^{1/2}_{x}g^{kl}_{x}\partial_{kx}\chi_{x}\partial_{lx}\chi_{x}    \right )g^{ij}_{x}\notag\\
 &+\frac{1}{2}\int D\chi \rho \, g_{x}^{1/2}\partial_{x}^{i}\chi_{x}\partial^{j}_{x}\chi_{x}~,\label{6 Matter source b}
\end{align}
where $\tilde{H}_{\perp x}^{0}$ was given in eq.(\ref{6 e-H b}), and where $\partial^{i}_{x}\chi_{x} = g^{ij}_{x}\partial_{jx}\chi_{x}$.

In total, the equations of motion for the gravitational field follow Hamilton's equations for the gravitational variables $(g_{ijx},\pi^{ij}_{x})$, given by eq.(\ref{6 Metric time deriv b}) for the evolution of the metric, and for the conjugate momentum $\pi^{ij}_{x}$ we have
\begin{eqnarray}
\partial_{t}\pi^{ij}_{x} - \pounds_{n}^{G}\pi^{ij}_{x} - \pounds_{N^{i}}\pi^{ij}_{x} = -N_{xt}\frac{\partial \tilde{H}_{\perp x}}{\partial g_{ijx}} ~.\label{6 Pi time deriv b}
\end{eqnarray}
This is an equation in which geometrical variables on the left-hand side are sourced by the variables $(\rho,\Phi)$, which contain all the information available about the field $\chi_{x}$ on the right-hand side. Below we will discuss this equation in the presence of ``quantum matter."

\section{Quantum sources of gravitation}\label{ED_QT}
The transition to what may be termed a \emph{quantum} form of dynamics amounts to an appropriate choice of the functional $F_{x}[\rho;g_{ij}]$ (see e.g., \cite{Ipek Caticha 2014}\cite{Ipek et al 2017}\cite{Ipek et al 2019}). In particular, for $F_{x}[\rho;g_{ij}]$ we choose exactly the local quantum potential introduced as part of (\ref{e Hamiltonian final}). A convenient choice for the coupling constant is $\lambda = 1/8$.\footnote{As argued, for instance, in \cite{Bartolomeo et al 2014}, there is no loss of generality in making this choice. For the case of nonrelativistic particles it can be proved \cite{Caticha 2019c} that an ED that preserves the appropriate symplectic and metric structures implies the presence of a quantum potential with the correct coefficient.} (This choice of $\lambda$ makes plain that we work with a system of units where $\hbar = c = 1$.)

The connection to conventional quantum theory is made explicit by a change of variables from the probability $\rho$ and phase $\Phi$ to the complex variables
\begin{eqnarray}
\Psi= \rho^{1/2}e^{i\Phi}\quad\text{and}\quad \Psi^{*} = \rho^{1/2}e^{-i\Phi}~.\label{6 Define Psi}
\end{eqnarray}
Such a change of variables is, in fact, a canonical transformation, and so, the new variables form a canonical pair given by $(\Psi, i\Psi^{*})$, which obey a natural generalization of the standard Poisson bracket relations
\begin{eqnarray}
\left \{ \Psi[\chi], i\Psi^{*}[\chi^{\prime}]  \right \} = \delta[\chi - \chi^{\prime}]~,
\end{eqnarray}
where $ \delta[\chi - \chi^{\prime}]$ is a Dirac delta \emph{functional}.

\subsection*{Quantum operators and geometrodynamics revisited}
Having chosen an $F_{x}[\rho;g_{ij}]$ of the type described above, the ensemble generators $\tilde{H}_{Ax}$ take a particularly special form
\begin{eqnarray}
\tilde{H}_{Ax} = \int D\chi \Psi^{*}\hat{H}_{Ax}\Psi = \left \langle \hat{H}_{Ax}  \right \rangle~, \label{6 Quantum Super Hamiltonian ensemble}
\end{eqnarray}
which can be viewed as both the expected value with respect to the probability $\rho$ of a Hamiltonian density, as in (\ref{e Hamiltonian final}), as well as the expectation value of the local Hamiltonian operators
\begin{subequations}
\begin{eqnarray}
\hat{H}_{\perp x} &=& - \frac{1}{2g^{1/2}}\frac{\delta^{2}}{\delta\chi^{2}_{x}}+\frac{g^{1/2}}{2}g^{ij}\partial_{i}\chi_{x}\partial_{j}\chi_{x}+g^{1/2}V_{x}(\chi_{x};g_{ij}) \label{6 Quantum Super Hamiltonian operator} \\
\hat{H}_{i x} &=& i\,\partial_{i}\chi_{x}\frac{\delta}{\delta\chi_{x}}~, \label{6 Quantum Super momentum operator}
\end{eqnarray}
\end{subequations}
with respect to the quantum state $\Psi$. To obtain the ``matter" contribution in eq.(\ref{6 Pi time deriv b}) for the conjugate momentum $\pi^{ij}_{x}$, note that the metric appears in $\tilde{H}_{\perp x}$ through the density $g_{x}^{1/2}$ and the inverse metric $g^{ij}_{x}$. Variations of these quantities with respect to the metric $g_{ijx}$ are given by \cite{Carroll 2004}
\begin{eqnarray}
\delta g_{x} = g  \,g^{ij}_{x} \, \delta g_{ijx}\quad\text{and}\quad \delta g^{ij}_{x} = g^{ik}g^{jl}\delta g_{kl}~.\notag
\end{eqnarray}
Then
\begin{eqnarray}
  \frac{\partial \tilde{H}_{\perp x}}{\partial g_{ijx}} = \int D\chi \, \Psi^{*}  \frac{\partial \hat{H}_{\perp x}}{\partial g_{ijx}} \Psi~,\notag
\end{eqnarray}  
where we have defined the operator
\begin{eqnarray}
  \frac{\partial \hat{H}_{\perp x}}{\partial g_{ijx}} = \partial^{i}_{x}\chi_{x}\partial^{j}_{x}\chi_{x}+g^{ij}\left (\frac{1}{2g_{x}^{1/2}}\frac{\delta^{2}}{\delta\chi_{x}^{2}}+\frac{g^{1/2}_{x}}{2}g^{kl}\partial_{k}\chi_{x}\partial_{l}\chi_{x}+V_{x}(\chi_{x})\right )~.\label{6 Quantum stress operator}
\end{eqnarray}

\paragraph*{Geometrodynamics with quantum sources--- }
Our goal here is to rewrite the main equations of geometrodynamics with sources given by quantum matter. We begin first with the constraint equations. With the aid of the local operators introduced in (\ref{6 Quantum Super Hamiltonian operator}) and (\ref{6 Quantum Super momentum operator}), the total Hamiltonian generators take the explicit form
\begin{subequations}
\begin{eqnarray}
H_{\perp x} &=& \frac{\kappa}{g_{x}^{1/2}}\left (2\pi^{ij}_{x}\pi_{ijx} -\pi^{2}_{x}\right) - \frac{g^{1/2}_{x}}{2\kappa}R_{x} + \int D\chi \, \Psi^{*} \hat{H}_{\perp x} \Psi~ \label{6 Quantum Hamiltonian}\\
H_{i x} &=& -2\nabla_{kx}\left (\pi^{kj}_{x}g_{ijx}\right )+ \int D\chi \, \Psi^{*} \hat{H}_{i x} \Psi ~,\label{6 Quantum momentum}
\end{eqnarray}
\end{subequations}
which are, of course, subject to the constraints
\begin{eqnarray}
H_{\perp x}\approx 0\quad \text{and}\quad H_{i x}\approx 0~. \label{Quantum constraints}
\end{eqnarray}
As relations (\ref{6 Quantum Hamiltonian}) and (\ref{6 Quantum momentum}) coupled together with (\ref{Quantum constraints}) make abundantly clear, the quantum state and the geometrical variables can no longer be treated as independent. This has important consequences for the time evolution of $\Psi$.

Moving on, note that the dynamical equation for the metric, given in eq.(\ref{6 Metric time deriv b}), does not depend directly on the choice of $F_{x}$ and therefore remains unchanged in the quantum context. The dynamical equation for the conjugate momentum $\pi^{ij}_{x}$, however, is modified. Using the operator introduced in (\ref{6 Quantum stress operator}), this becomes
\begin{eqnarray}
\partial_{t}\pi^{ij}_{x} - \pounds_{n}^{G}\pi^{ij}_{x} - \pounds_{N^{i}}\pi^{ij}_{x} = -N_{xt}\int D\chi \, \Psi^{*}\frac{\partial \hat{H}_{\perp x}}{\partial g_{ijx}}\Psi~,\label{6 Quantum Pi evolution}
\end{eqnarray}
where $\pounds_{m}^{G}\pi^{ij}_{x} = $ and $\pounds_{N^{i}}\pi^{ij}_{x}$ are given in eqns.(\ref{6 Pi Lie normal G}) and (\ref{6 Pi Lie tangential G b}), respectively. This is the crucial equation in which the dynamical geometry is itself affected by the epistemic state $\Psi$.

Putting it all together, the eqns.(\ref{6 Quantum Hamiltonian})-(\ref{6 Quantum Pi evolution}), and eq.(\ref{6 Metric time deriv b}) for the metric, constitute a system of equations that are \emph{formally} equivalent to the SCEE put in the canonical form. However, that is where the similarities end. On several key issues of interpretation, in particular, the ED approach is vastly different from the SCEE as they are normally understood. Far from being trivial, these distinctions turn out to be quite important since many objections (see e.g., \cite{Kibble et al 1980}\cite{Unruh 1984}) to the usual SCEE are, in fact, based on such considerations.

\subsection*{Quantum dynamics}
One advantage of the complex variables $(\Psi,\Psi^{*})$ is that the dynamics takes a \emph{familiar} form. Indeed, since $\Psi$ and $\Psi^{*}$ are just functions of our canonical variables $(\rho,\Phi)$ we can just use eq.(\ref{6 Def time derivative c}) to determine their evolution along a time parameter $t$, which gives
\begin{eqnarray}
\partial_{t}\Psi_{t}[\chi] =  \int dx \left \{\Psi_{t}[\chi],H_{Ax}\right \}N^{A}_{xt}~.\label{6 Quantum Evolution Psi a}
\end{eqnarray}

The tangential component is obtained in a straightforward fashion by
\begin{eqnarray}
 \left \{\Psi_{t}[\chi],H_{ix}\right \} = \partial_{i}\chi_{x}\frac{\delta\Psi_{t}[\chi]}{\delta\chi_{x}}~,\label{6 Quantum Tangential Psi}
\end{eqnarray}
which is reasonable since this is just the Lie derivative of $\Psi[\chi]$ along the surface.\footnote{Compare, for example, to eqns.(\ref{6 Tangential deformation rho c}) and (\ref{6 Tangential deformation Phi a}) for $\rho[\chi]$ and $\Phi[\chi]$.} The local normal evolution of $\Psi$, on the other hand, is given by
\begin{eqnarray}
 \left \{\Psi_{t}[\chi],\tilde{H}_{\perp x}\right \} =-i \, \hat{H}_{\perp x}\Psi_{t}[\chi]~.
\end{eqnarray}
Inserting these results into eq.(\ref{6 Quantum Evolution Psi a}) for a general evolution of $\Psi_{t}[\chi]$, we then have
\begin{eqnarray}
i \, \partial_{t}\Psi_{t}[\chi] = \int dx\left (N_{xt}\hat{H}_{\perp x}+N^{i}_{xt}\hat{H}_{ix}\right )\Psi_{t}[\chi]~.\label{6 Quantum Schrodinger equation}
\end{eqnarray}
Finally, substituting eqns.(\ref{6 Quantum Super Hamiltonian operator}) and (\ref{6 Quantum Super momentum operator}) in for $\hat{H}_{\perp x}$ and $\hat{H}_{i x}$, respectively, yields the equation
\begin{eqnarray}
i \hbar \partial_{t}\Psi_{t} &=& \int dx N_{xt}\left(-\frac{\hbar^{2}}{2g^{1/2}}\frac{\delta^{2}}{\delta\chi^{2}_{x}}+\frac{c^{2}}{2}g^{1/2}_{x}\, g^{ij}\partial_{i}\chi_{x}\partial_{j}\chi_{x}+g^{1/2}V_{x}\right )\notag\\
&-& \hbar c \,\int dx N^{i}_{xt}\partial_{i}\chi_{x}\frac{\delta}{\delta\chi_{x}}\Psi_{t}~,\label{Quantum Schrodinger equation hbar c}
\end{eqnarray}
where we have reinstated the constants $\hbar$ and $c$, as appropriate. Here equation (\ref{Quantum Schrodinger equation hbar c}) is ostensibly just a linear equation for the complex variable $\Psi_{t}$, which \emph{suggests} calling it a Schr\"{o}dinger functional equation. We discuss below.

\paragraph*{But is it quantum?--- }
The coupling to classical gravity described above implies violations of the superposition principle. To see this, we note that geometrodynamics is a constrained dynamical system, where the operative equations are given by the Hamiltonian constraints in (103). Solving these constraints often involves determining the components of the metric in terms of the quantum sources, which then gets fed back into evolution equations for $\Psi$.\footnote{This dynamic can most readily be seen within the linearized gravity regime, or the weak field limit. In fact, there are some arguments (see e.g., \cite{Bahrami et al 2014}) that the so-called Newton-Schr\"{o}dinger equation, which is a non-linear equation, results from the Newtonian limit of the semi-classical Einstein equations. Investigation of this result coming from ED is underway.} This feedback leads to a non-linear time evolution (see e.g., \cite{Kibble et al 1980}) in which the $\Psi$ itself appears as a potential in eq.(\ref{Quantum Schrodinger equation hbar c}).

A natural question that arises is whether such a dynamics can rightfully be called ``quantum" or not. But this line of inquiry is rather misguided because the search for QG is, in many respects, the search for which criteria, in fact, constitute a quantum theory in the first place. Since the ED model formulated here involves the presence of $\hbar$, a wave equation for a complex wave function, an uncertainty principle and non-local correlations, and limits to the standard quantum formalism in a flat space-time, we claim that it is \emph{these} ingredients that are fundamental for defining a quantum theory. Additional features, such as the superposition principle, emerge here only as an effect confined to a limiting regime.

\section{Concluding remarks}
\label{conclusion}

The ED developed here couples quantum ``matter" to classical gravity on the basis of three key principles: (1) A properly entropic setting wherein the dynamics of probability is driven by information encoded into \emph{constraints}. (2) The preservation of a symplectic structure as a primary criterion for updating the evolving constraints. (3) Foliation invariance symmetry, enacted by representing the DHKT ``algebra" in terms of the relevant Poisson brackets.

Our approach results in several interesting features.
Although written in the relatively less common language of geometrodynamics,
the eqns.(\ref{6 Metric time deriv b}) and
(\ref{Quantum constraints})-(\ref{6 Quantum Pi evolution}) are
formally equivalent to the so-called semi-classical\ Einstein equations
(SCEE)
\begin{eqnarray}
G_{\mu\nu}=8\pi G\langle\hat{T}_{\mu\nu}\rangle
~,\label{6 Semi classical equations}%
\end{eqnarray}
with classical Einstein tensor $G_{\mu\nu}$ (see e.g., \cite{Carroll 2004}),
but sourced by the expected value of the quantum stress-energy
tensor.\footnote{The components of the quantum stress-energy tensor are
related in a simple manner to the ensemble quantities in
eqns.(\ref{6 Quantum Super Hamiltonian ensemble}) and
(\ref{6 Quantum stress operator}). The derivation of the SCEE from an action
by Kibble and Randjbar-Daemi in \cite{Kibble et al 1980} makes this
relationship more precise.} Such a theory of gravity has long been seen as a
desirable step intermediate to a full theory of QG, in part because it
contains well-established physics --- QFTCS and classical GR
--- in the limiting cases where they are valid. But there has been much debate
(see e.g., \cite{Unruh 1984}\cite{Eppley Hannah 1977}\cite{Page Geilker 1981}\cite{Duff 1981}), on the other hand, as to the status of
semi-classical theories as true QG candidate; with many harboring a negative view.

Here we do not propose a definitive rebuttal to those critics, but note that
the ED formulation of SCEE has certain features that allow it to evade the
most cogent criticisms. For one, a problem that is often raised against the
SCEE is that it is proposed in a rather \emph{ad hoc} manner, based on
heuristic arguments. Indeed, the usual argument for the expected value on the
right hand side of eq.(\ref{6 Semi classical equations}) is at best a guess.
In ED, on the other hand, the coupling of geometry to the expected value of
quantum operators --- made explicit in
eqns.(\ref{6 Quantum Hamiltonian})-(\ref{6 Quantum Pi evolution})
--- is \emph{derived} on the basis of well-defined assumptions and
constraints. In other words, we derive the SCEE in ED from \emph{first
principles}. Indeed, these principles have already been tested elsewhere: not
only do they provide a reconstruction of GR through the work of DHKT, but they
also provide a reconstruction of both nonrelativistic quantum mechanics and relativistic quantum field theory. Furthermore, while there is no guarantee
that such constraints and assumptions are adequate for a full QG theory, ED
provides a framework wherein additional information can be incorporated as needed.

One source of confusion with the SCEE is the issue of what happens when the
location of a macroscopic source is uncertain. (See \emph{e.g.}, \cite{Unruh 1984}\cite{Bahrami et al 2014}.) Consider, for example, a
macroscopic mass $m$ that is equally likely to be at $x_{1}$ or at $x_{2}$.
Will the gravitational field itself be equally likely to point either towards
$x_{1}$ or towards $x_{2}$? Or, will the gravitational field be as if
generated by a mass $m$ located at the expected position $(x_{1}+x_{2})/2$?
The ED resolution of this paradox shows the advantage of having a derivation
from first principles. The~$\langle\hat{T}_{\mu\nu}\rangle$ on the right of
the SCEE is not the expectation value taken over any arbitrary source of
uncertainty. The $\langle\hat{T}_{\mu\nu}\rangle$ was derived, or better, it
was inferred from a very specific type of information that leads, by an
abuse of language, to what one might call a pure quantum state. The issue then is what
is the gravitational field generated by a pure state that happens to be a
macroscopic "Schr\"{o}dinger cat"? To the extent that this is a
\textquotedblleft pure\textquotedblright\ state then ED gives a sharp
prediction: the gravitational field is generated by an $\langle\hat{T}_{\mu
\nu}\rangle$ centered at the average position. But such a Schr\"{o}dinger cat
state cannot be physically realized: it would immediately suffer
decoherence. To analyze such a situation the ED framework would need to be
extended to incorporate information (\emph{i.e.}, additional constraints) that
describes additional sources of uncertainty. In this extended ED it is
conceivable that the gravitational field itself would be uncertain. So the
conclusion is that there is no paradox; the two different predictions
correspond to two different inferences arising from two different pieces of
information.  

Another family of objections revolve around paradoxes related to the problem
of quantum measurement. As pointed out long ago by Kibble \cite{Kibble 1978} these
criticisms are premature as long as the interpretation of quantum mechanics is
problematic and the problem of the collapse of the wave function remains unsolved. The issue is whether a measurement that induces a collapse of the
wave function would result in a discontinuous change in the expected
$\langle\hat{T}_{\mu\nu}\rangle$ with its attendant violations of causality.  

%Another family of objections revolve around paradoxes related to the problem
%of quantum measurement. As pointed out long ago by Kibble \cite{Kibble 1978} these
%criticisms are premature as long as the interpretation of quantum mechanics is
%problematic and there is not solution to the problem of collapse of the wave
%function. The issue is whether a measurement that induces a collapse of the
%wave function would result in a discontinuous change in the expected
%$\langle\hat{T}_{\mu\nu}\rangle$ with its attendant violations of causality.  

Within the ED approach such objections do not arise. ED solves the problem of
interpretation by starting with a clear definition of the ontology, which
eliminates the interpretation problem. Furthermore, by including in its very
foundation the epistemic tools for inference, ED also solves the problem of
measurement \cite{Caticha Johnson 2012}\cite{Caticha Vanslette 2017}. In ED a measument
device is not described as a black box subject to rules that violate the
Schr\"{o}dinger equation but as a physical process subject to the very same
laws that describe the rest of the world. As a result in a fully covariant ED
such as the model developed in this paper the process of measurement is
described by the same causal flow of probability that characterizes any other
physical process.

Yet another argument that has been raised against the SCEE is that the left
hand side, featuring the gravitational field, is a \textquotedblleft
physical\textquotedblright\ ontic field, while the right hand side contains
the quantum state $\Psi$, which is epistemic. Within the ED approach, however, the fact that $\langle \hat{T}_{\mu\nu}\rangle$ on the right hand side of (\ref{6 Semi classical equations}) is epistemic necessarily forces the left hand side, the geometry, to be epistemic too.\footnote{Of course, other theories will lead to other interpretations. If one were to adopt alternative non-Bayesian interpretations of probability (say, frequentist or propensity) or of QM (e.g., a Bohmian ontic wave function), then the geometry of spacetime would not need to be epistemic.} Indeed, a specific proposal along these lines has been offered in  \cite{Caticha 2016}\cite{Caticha 2019b}, suggesting that the geometry of space-time itself can be given an entropic underpinning.%\footnote{Here we adopt the view that entropy is a \emph{tool} for ranking epistemic probability distributions (see e.g., \cite{Caticha 2012}). Therefore any quantity which is entropic must also be epistemic.}

Finally, the Schr\"{o}dinger equation derived here is quite unorthodox in that
the dynamics of $\Psi$ follows a \emph{non-linear} equation. This is quite
problematic in the standard view of QT, where linearity is held as sacrosanct.
But, as mentioned in the introduction, in the ED approach to quantum theory
Hilbert spaces are not fundamental; they are introduced as a convenient trick
precisely because of the calculational advantage of the linearity they induce.
In the model developed here such a trick cannot be carried out and the
superposition principle becomes the first casualty in a successful coupling of
quantum matter with dynamical gravity. 

The nonlinearity of the SCEEs does extreme violence to our understanding of
quantum theory and raises many questions. Is the introduction of
Hilbert spaces at all justified? Are density matrices at all useful? Can we
expect something like the evolution from pure to mixed states? Or, what seems
more likely, the very concepts of pure and mixed states are so linked to the
concept of Hilbert spaces that, absent the latter, the more useful description
is just in terms of probabilities $\rho$ and phase fields $\Phi$. What is the
generalization of von Neumann's entropy for such states?  Can non-orthogonal
states be distinguished? Can quantum states be cloned? The point of even
raising such questions is precisely to emphasize the very different research
directions one is led to once the standard tools of quantum mechanics are no
longer fundamental and/or available. We must be open to the possibility that
the proper way to model gravity might be as neither a quantum nor a classical
theory but something else altogether --- perhaps an inferential model based on
information geometry in the spirit of the ED approach to QM.

%%%%%%%%%%%%%%%%%%%%%%%%%%%%%%%%%%%%%%%%%%
\subsection*{Acknowledgments}
We would like to thank M. Abedi, D. Bartolomeo, N. Carrara, N. Caticha, F. Costa, S. DiFranzo, K. Knuth, S. Nawaz, P. Pessoa, and K. Vanslette for many valuable discussions on entropy, inference, quantum theory, and much more. We would also like to thank the physics department of University at Albany---SUNY for their continued support.

\end{document}